\documentclass[%
 reprint,
 amsmath,amssymb,
 aps,
notitlepage,
]{revtex4-2}

\usepackage{graphicx}
\usepackage{dcolumn}
\usepackage{bm}

\usepackage{url}
\usepackage[hyperindex,breaklinks]{hyperref}
\hypersetup{linkcolor=blue, citecolor=red, colorlinks=true}

\usepackage{placeins} 
\usepackage{xcolor}

\usepackage{xparse}

\NewDocumentCommand{\tens}{t_}
 {%
  \IfBooleanTF{#1}
   {\tensop}
   {\otimes}%
 }
\NewDocumentCommand{\Log}{o}{%
  \IfNoValueTF{#1}{}{{}^{#1}\!}\log}%

\newcommand{\ket}[1]{| #1 \rangle}

\begin{document}

\preprint{APS/123-QED}

\title{Quantum walks with sequential aperiodic jumps}

\author{Marcelo A. Pires$^{1}$}
\thanks{piresma@cbpf.br}

\author{S\'{\i}lvio M. \surname{Duarte~Queir\'{o}s}$^{1,2}$}
\thanks{sdqueiro@cbpf.br}

\affiliation{
$^{1}$Centro Brasileiro de Pesquisas F\'isicas, Rio de Janeiro/RJ, Brazil
\\
$^{2}$National Institute of Science and Technology for Complex Systems, Brazil}

\date{\today}

\begin{abstract}
We analyze a set of discrete-time quantum walks for which the displacements on a chain follow binary aperiodic jumps according to three paradigmatic sequences: Fibonacci, Thue-Morse and Rudin-Shapiro. We use a generalized Hadamard coin, $\widehat C_{H}$, as well as a generalized Fourier coin,  $\widehat C_{K}$. We verify the QW experiences a slowdown of the wavepacket spreading --- $\sigma ^2 (t) \sim t^\alpha $ --- by the aperiodic jumps whose exponent, $\alpha$, depends on the type of aperiodicity. Additional aperiodicity-induced effects also emerge, namely: (i) while the superdiffusive regime ($1<\alpha<2$) is predominant, $\alpha$ displays an unusual sensibility with the type of coin operator where the more pronounced differences emerge for the Rudin-Shapiro and random protocols; (ii) even though the angle $\theta$  of the coin operator is homogeneous in space and time, there is a nonmonotonic dependence of  $\alpha$ with $\theta$. Fingerprints of the aperiodicity in the hoppings are also found when 
distributional measures such as the Shannon and von Neumann entropies, the Inverse Participation Ratio, the Jensen-Shannon dissimilarity, and the kurtosis are computed,  which allow assessing informational and delocalization features arising from these protocols and understanding the impact of linear and non-linear correlations of the jump sequence in a quantum walk as well. Finally, we argue the spin-lattice entanglement is enhanced by aperiodic jumps.
\end{abstract}

\keywords{Aperiodicity, novel difference between classical and quantum walks}
                              
\maketitle

\section{\label{sec:intro}Introduction}

Since its introduction, Quantum Walks(QWs)~\cite{aharonov1993quantum} have been understood as a means for comprehending ubiquitous complex dynamics ruled by quantum fields and mathematically described by a sequence of local (and unitary) operations that act on the quantum particle --- i.e., a cell occupied by a quantum particle --- and its internal degrees of freedom as well~\cite{kempe2003quantum,venegas2012quantum}. 
Among the instances which have profited from this approach we refer to problems within
algorithmics~\cite{ambainis2003quantum,kendon2006random,portugal2013quantum}, machine learning~\cite{paparo2014quantum} and experimental implementations~\cite{wang2013physical,neves2018photonic},
just to mention a few.

Still considering the scope of QWs, a relevant field of research has to do with quantum systems under high noise and randomness~\cite{kendon2003decoherence} where --- as occurs for classical systems --- nondeterministic elements are aimed at depicting some sort of interaction between the system and the environment~\cite{attal2012open,uchiyama2018environmental}.
Complementarily to different types of randomness~\cite{zeng2017discrete,vieira2013dynamically,di2018elephant,pires2019multiple}
either in the phase of the unitary transforms~\cite{zeng2017discrete,vieira2013dynamically} or the jump distribution~\cite{di2018elephant,pires2019multiple}, it is possible to assess the existence of sequencing in the protocol. The purpose of the present work is precisely to understand to what degree the existence of the aperiodic sequencing features impacts in the quantum statistical and informational properties of a quantum walk with such traits. To that, we consider three paradigmatic aperiodic sequences which strongly relate to quantum systems: Fibonacci~\cite{vaezi2014fibonacci}, Thue-Morse~\cite{doria1989thue} and Rudin-Shapiro~\cite{trabelsi2016narrow}.
\textcolor{black}{Explicitly, in employing those 
aperiodic sequences, we are able to gauge the impact of relevant and wide-ranging types of inhomogeneity wherewith it is possible to manipulate the (de)localization properties of a quantum system, which is a very handy tool namely in their applications like quantum algorithms and other protocols. Moreover, owing to the fact that these sequences have different degrees of linear and/or non-linear 
self-dependencies, the present work allows understanding the impact of non-linear correlations in delocalization phenomena and spin-lattice entanglement as well.}

The paper is organized as follows: in Sec.~\ref{sec:lit-rev} we put our work within the context of quantum walks subjected to noise and disorder by briefly reviewing the literature on this matter; in Sec.~\ref{sec:model}, we introduce our model and each protocol; in Sec.~\ref{sec:results}, we present the results for each aperiodic sequence case and in Sec.~\ref{sec:remarks} we address our final remarks on this research as well as setting forth an outlook for future steps.

\section{\label{sec:lit-rev}Literature review}

In its canonical version~\cite{aharonov1993quantum}, every step of a QW has the same size $J_t=1$.  The breakdown of such homogeneity paves the way to a set of phenomena such as multi-peaked distributions \cite{zhao2015one,ahmad2020randomizing}, localization~\cite{lavivcka2011quantum}, either inhibition~\cite{sen2019unusual,mukhopadhyay2019persistent,das2019inhibition} or hyperballistic spreading~\cite{di2018elephant} --- defined by the deviation of the wave packet --- $\sigma ^2 (t) \sim t^\alpha $ ---, enhancement of the spin-coin entanglement~\cite{sen2019scaling,mukhopadhyay2019persistent,pires2019multiple}. More recently, it was reported the emergence of multiple dynamical transitions~\cite{pires2019multiple}, especially, between ballistic ($\alpha = 2$) $\rightarrow$ diffusive ($\alpha = 1$) $\rightarrow$ superdiffusive ($1 < \alpha < 2$) $\rightarrow$  ballistic $\rightarrow$ hyperballistic regimes ($\alpha = 3$).
In all of those works time is discrete; complementarily, continuous-time QW with nonrandom position-dependent jumps have also been treated~\cite{mulken2008universal}. In the latter case, it was found a robust ballistic spreading for deterministic jumps following a power-law decaying step size. In Ref.~\cite{chattaraj2016effects}, it was shown the interplay between long-range hopping and long-range interaction breaks the symmetry of the two-particle correlation diagram. Last, open quantum L\'evy flights have been treated in the literature as well\cite{caceres2010quantum}.

The aforementioned studies with discrete-time QW share the feature of assuming random jump protocols. Herein, we address the problem of  QWs considering a deterministic protocol that is not periodic as well. Nonetheless, we specifically consider binary aperiodic sequences as the generator of the jumps performed by quantum particles on the chain. In spite of the fact that aperiodic sequences have been used as a source of disorder in the coin operator~\cite{ribeiro2004aperiodic,romanelli2009fibonacci,ampadu2012return,di2015massless,fillman2017resolvent,gullo2017dynamics,liu2018entanglement,andrade2018discrete}, this kind of protocol has not been embedded into the step operator. In this work, we fill that gap by letting the steps of the quantum walker follow one out of three paradigmatic aperiodic sequences, namely Fibonacci, Thue-Morse or Rudin-Shapiro, as previously mentioned. Besides the theoretical implications of our proposal stated in Sec.~\ref{sec:intro}, we can look at this work from an experimental perspective and point out the use of deterministic aperiodic disorder has the advantage of permitting very controllable dynamics~\cite{nguyen2019localized,nguyen2019quantum}.

\section{\label{sec:model}Model}

\subsection{Discrete-time quantum walk}

We consider a two-state quantum walker moving on $x \in \mathbb{Z}$ in a way that the wavefunction, at step $t \in \mathbb{N}$, is given by 
\begin{equation}
\Psi_{t} =  \sum_{x}
\left(   
\psi_{t}^{D} (x) \ket{\downarrow} +  \psi_{t}^{U} (x) \ket{\uparrow}
\right) \tens{} \ket{x},
\label{wavefunction}
\end{equation}
where $ \psi_{t}^{D} (x)$ and $\psi_{t}^{U} (x)$ are the  position and time dependent probability amplitudes associated with the two degrees of freedom $\{\downarrow,\uparrow\}$ of the quantum particle,  respectively. 
That is to say, the QW dynamics lives in a composite Hilbert space $\mathcal{H}_2 \otimes \mathcal{H}_\mathbb{Z}$. The evolution of $ \Psi_{t}$ is ruled by 
\begin{equation}
\Psi_{t+1}=  \hat{ W_{t} }  \Psi_t,
\label{eq:FDE0}
\end{equation}
where the operator 
\begin{equation}
\hat W_{t} \equiv \hat{T}(\hat{C}\otimes \hat {\mathcal{I}} _\mathbb{Z})
\end{equation}
is composed of two other operators, each acting on its respective Hilbert sub-space.
Accordingly, we have
\begin{itemize}

\item the coin operator:
\begin{equation}
\widehat{C} \equiv 
c_{11}
|\uparrow\rangle \langle\uparrow|
+
c_{12}
|\uparrow\rangle \langle\downarrow|
+
c_{21}
|\downarrow\rangle \langle\uparrow|
+
c_{22}
|\downarrow\rangle\langle\downarrow|,
\label{Eq:rotmat}
\end{equation}
with $c_{ij} $ standing for the matrix elements corresponding to the quantum coin operator, so that $ c_{12} $ and $ c_{21} $ appraise the coupled evolution of the components $\psi_{t}^{U} (x) $ and $\psi_{t}^{D} (x) $;

\item the spin-dependent hopping operator: 
  \begin{equation}
\widehat{T} \equiv  
|\downarrow\rangle \langle\downarrow|
\tens
\sum_x |x-J_t\rangle \langle x|
+
|\uparrow\rangle \langle\uparrow|
\tens
\sum_x |x+J_t\rangle \langle x|,
  \end{equation}
where $J_t$ is the step size, which will be discussed in detail shortly.

\end{itemize}

With respect to the quantum coin, we employ either the generalized Hadamard coin
$\widehat C_{H}$ or a generalized Fourier coin $\widehat C_{K}$ (also known as Kempe-like coin\cite{kempe2003quantum}): 
\begin{equation}
\widehat C_{H}
\equiv 
\begin{pmatrix}
\cos \theta &   \sin \theta \\
 \sin \theta  & -\cos \theta
\end{pmatrix},
\quad\quad
\widehat C_{K}
\equiv 
\begin{pmatrix}
\cos \theta &  i \sin \theta \\
i \sin \theta  & \cos \theta
\end{pmatrix}.
\end{equation}

We now define the initial condition as the localized state:
 \begin{equation}
 \Psi_{0}  = \frac{1}{\sqrt{2}} \delta_{x,0}  \left( \ket{\downarrow} + e^{i\phi}\ket{\uparrow} \right) \tens{} \ket{x}
 \end{equation}

 In order to have symmetric distributions we set $\phi=\pi/2$ for $\widehat C_{H} $ and 
$\phi=0$ for $\widehat C_{K}$ \cite{pires2019multiple}.

\subsection{Jump protocol}

\begin{figure}[t]
  \centering
\includegraphics[scale=0.38]{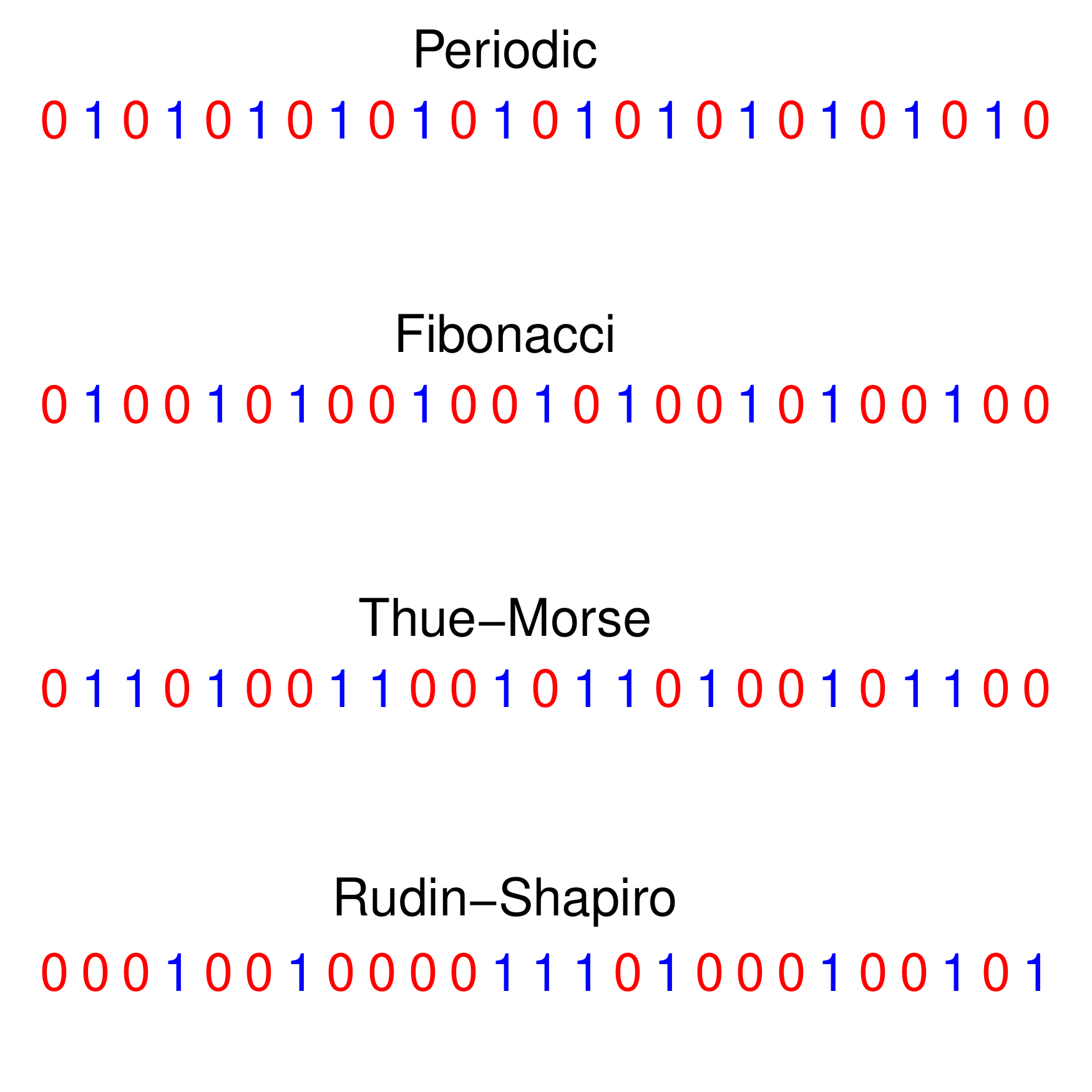}
\caption{Paradigmatic protocols to generate the jump sequence $J_t=1+b_t$: Periodic, Fibonacci, Thue-Morse, and Rudin-Shapiro. In addition, we have used the unbiased random case as well.  For ulterior comparison, we have considered the standard case where $b_t=0 \rightarrow J_t=1 \ \forall \ t$.}
\label{fig:bin-seq}
\end{figure}

\begin{figure*}[t]
          \centering
\includegraphics[scale=0.36]{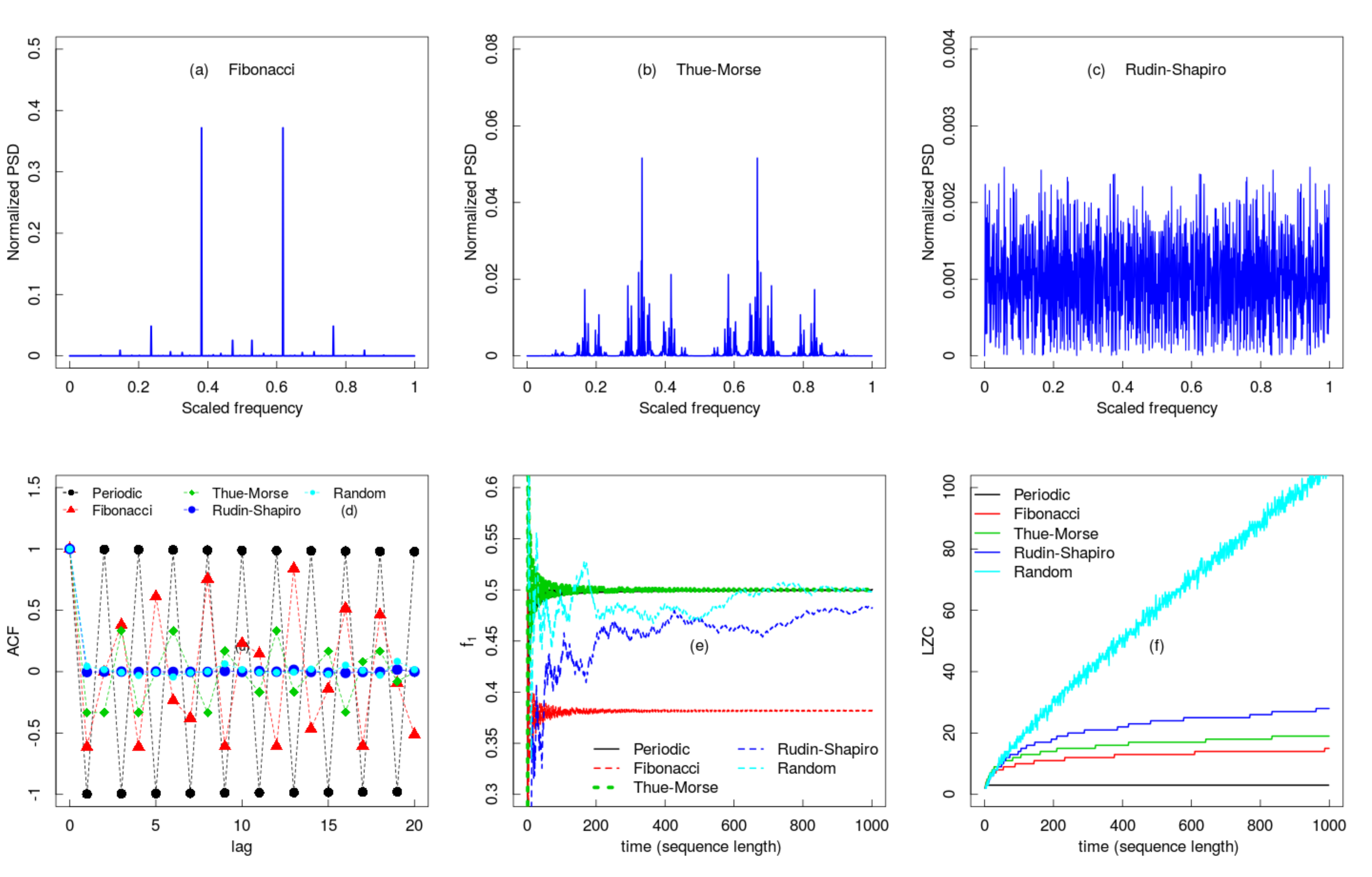}

\caption{Main properties of the aperiodic sequences Fibonacci, Thue-Morse, Rudin-Shapiro. (a-c) Normalized power spectrum density versus the scaled frequency $w/w_{\max}$. (d) Autocorrelation versus lags. (e) Fraction of 1 over time. (f) Lempel-ziv complexity over time. For comparison, we also show the corresponding results for the random and periodic sequence when needed.}
\label{fig:acf-psd-lzc}
\end{figure*}

\begin{figure*}[t]
     \centering
\includegraphics[scale=0.42]{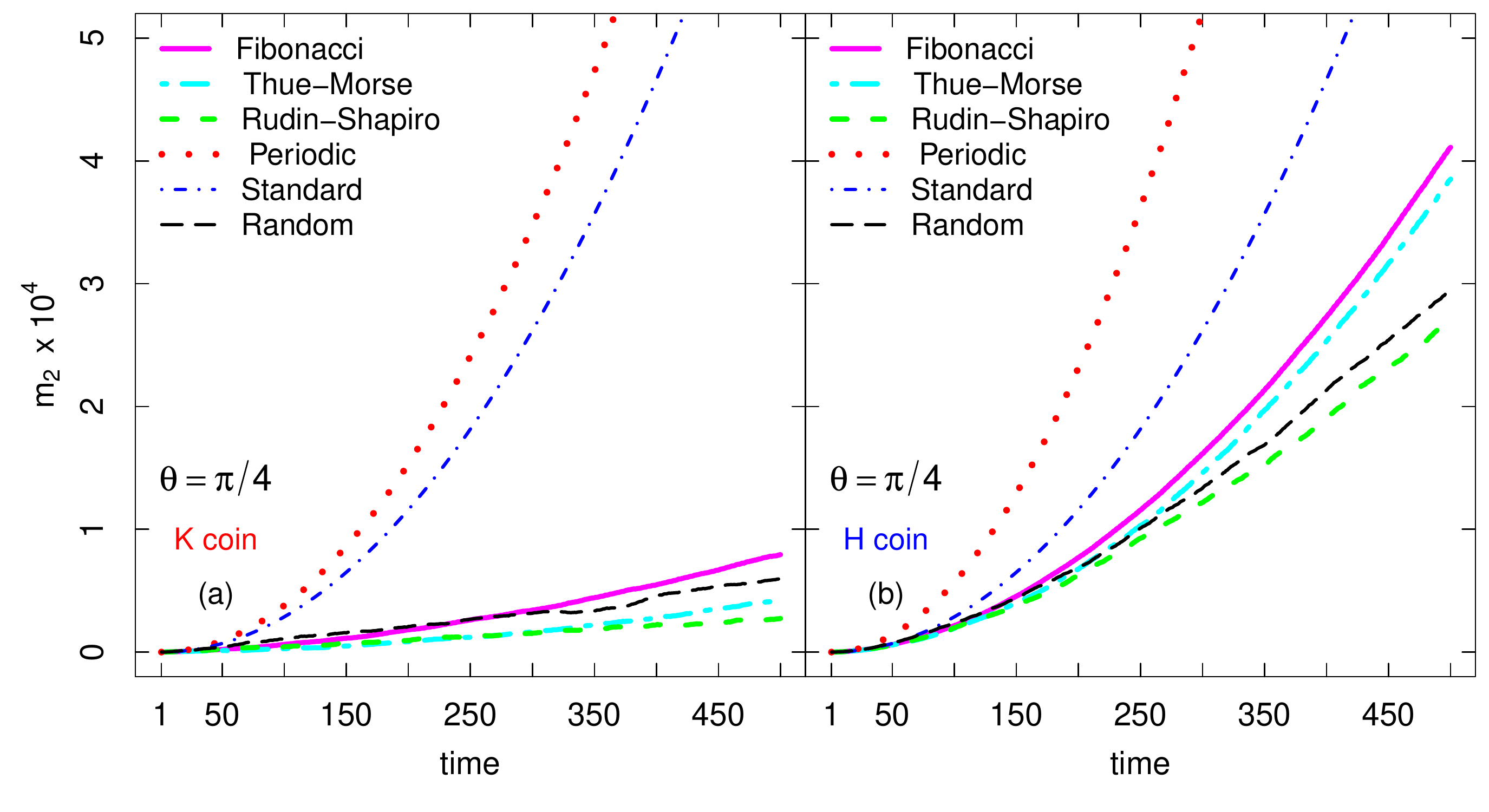}
\includegraphics[scale=0.42]{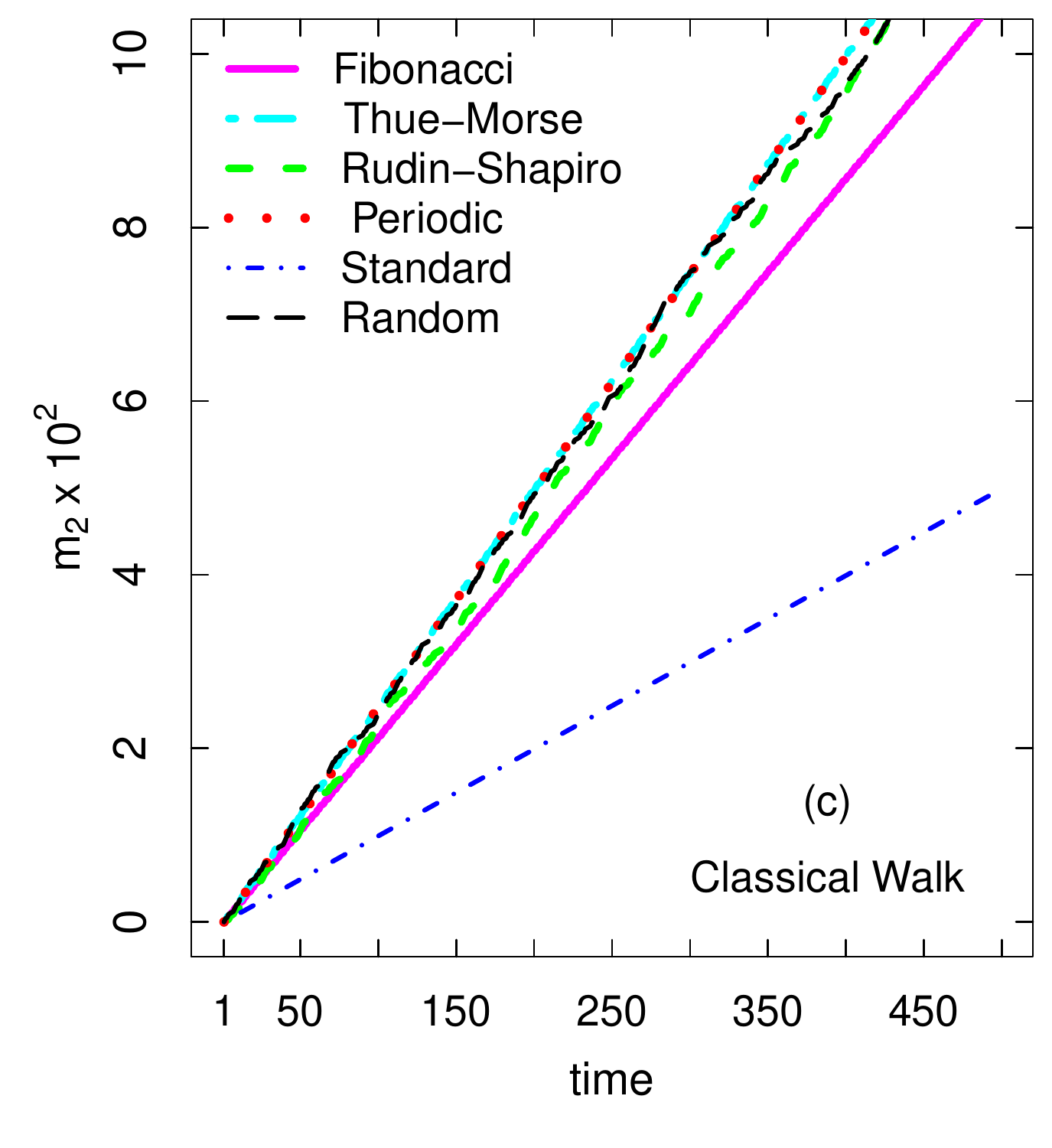}

\includegraphics[scale=0.42]{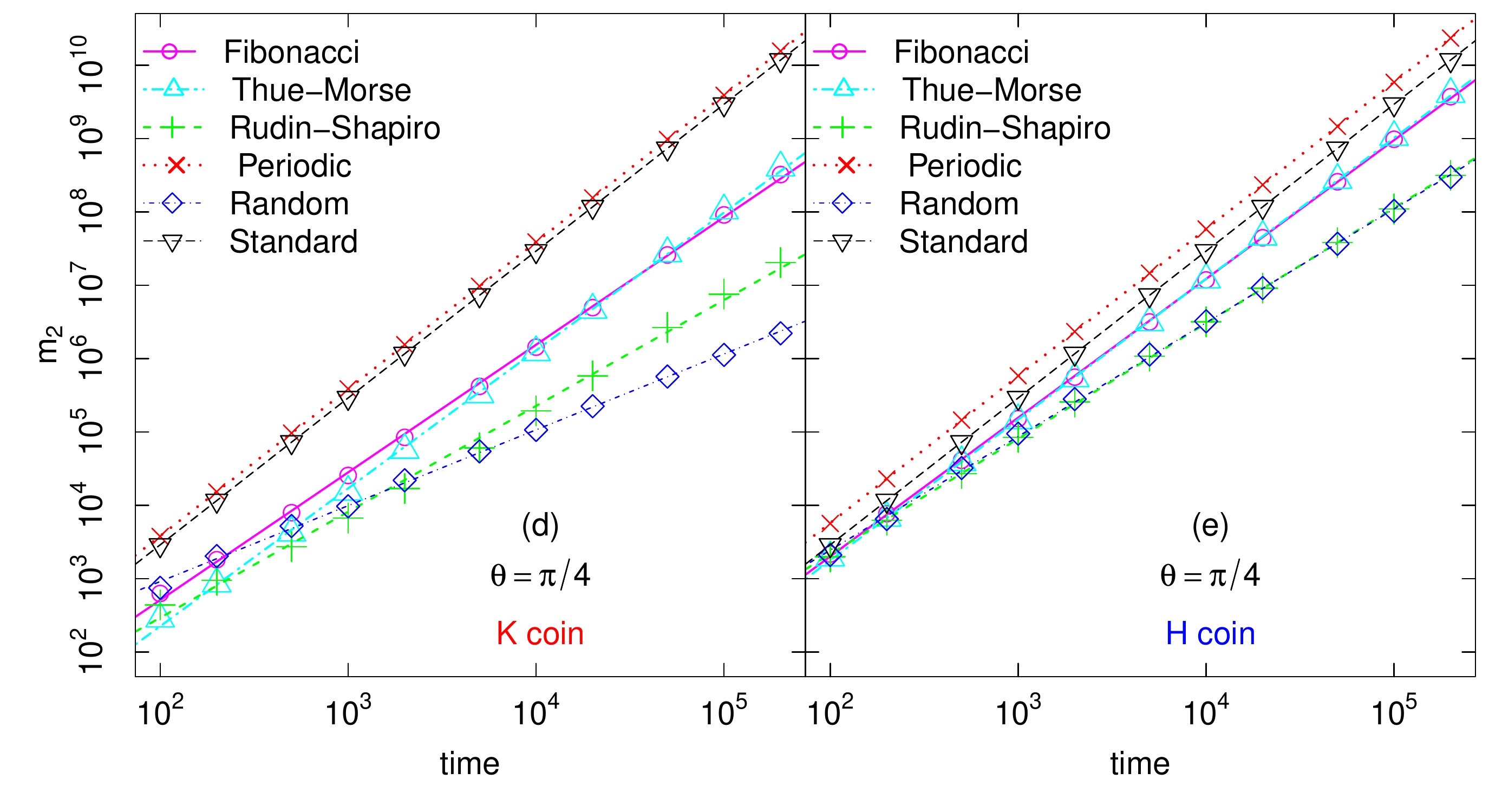}
\includegraphics[scale=0.42]{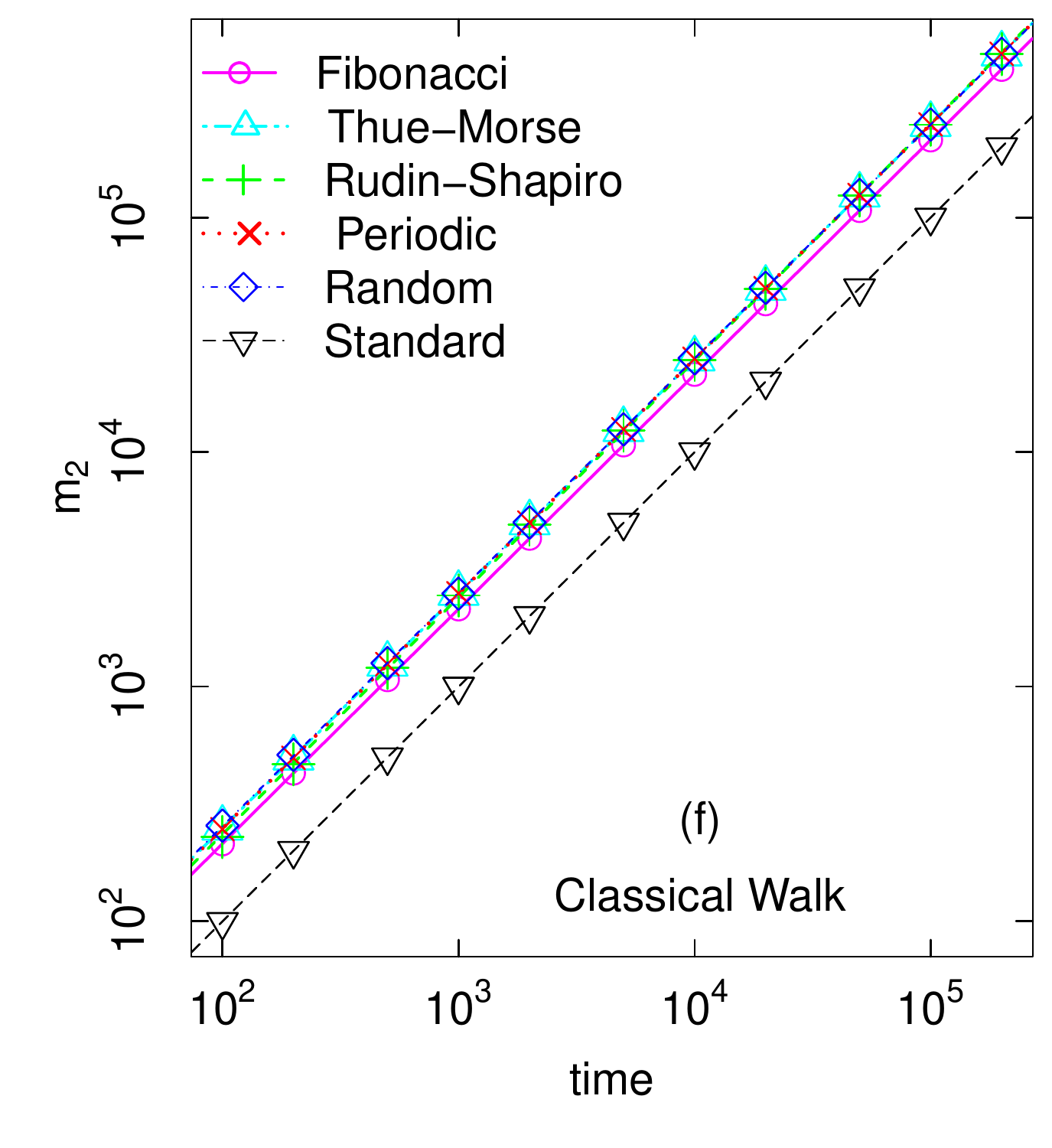}

\caption{(a-b) Spreading measure $m_2(t) = \overline{x^2}$ versus time for the quantum and classical walk with aperiodic, periodic and random protocols of jumps. (c-d) The symbols are the same as in previous panels but in $\log$-$\log$ scale and the lines represent the numerical adjustment $m_2(t) = \rm{const} \,\, t^\alpha$.
}
\label{fig:x2-qw-crw-log}
\end{figure*}

\begin{figure*}[t]
  \centering
\includegraphics[scale=0.42]{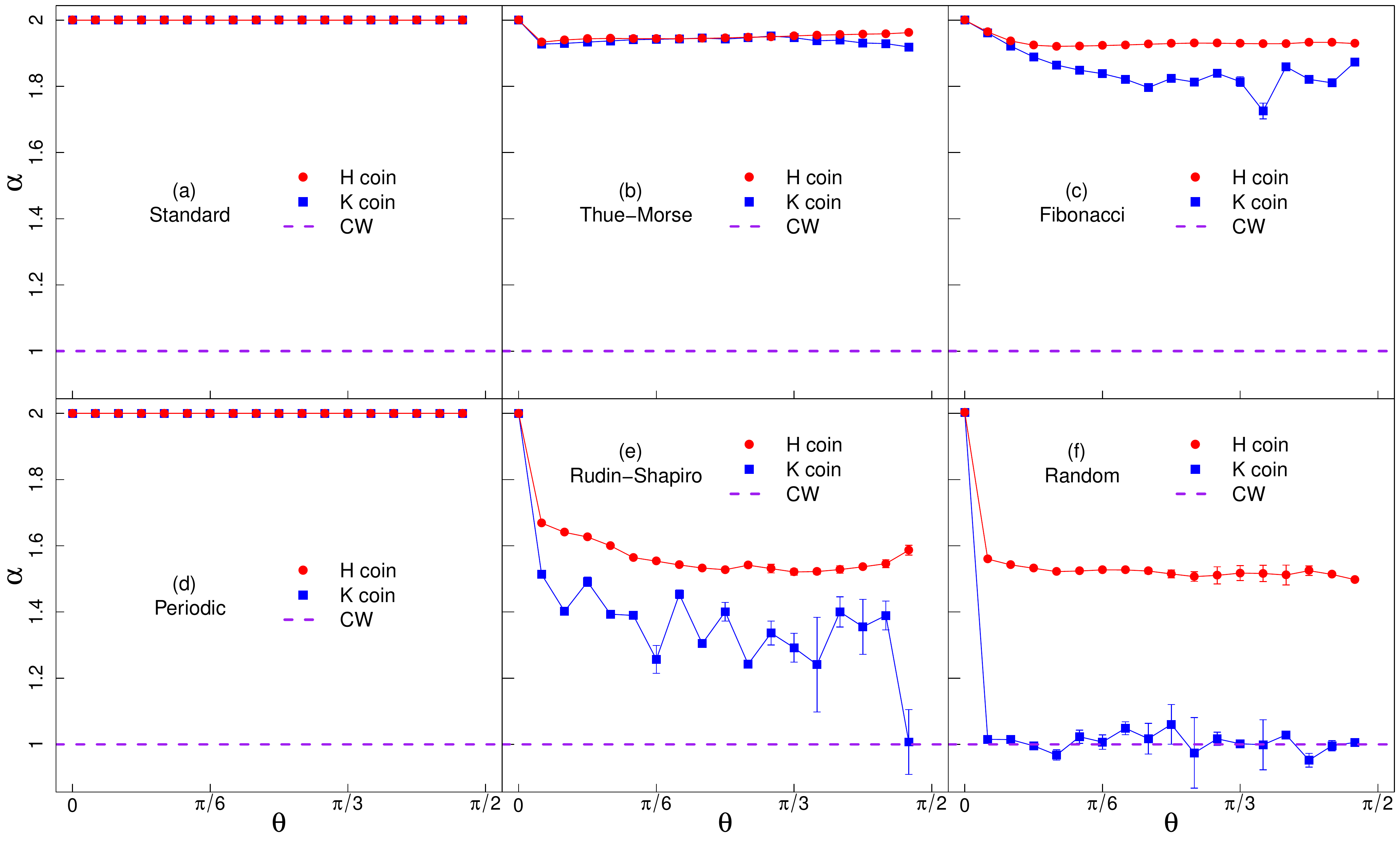} 
\caption{Dynamic regimes for the QW and CW under aperiodic/periodic chain of jumps.  We use $t_{max}=2\times 10^5$ to estimate $\alpha$ from $m_2  \sim t^\alpha$. Each point corresponds to an average over the two initial seeds of the binary sequence: $b_0=0$ ($J_0=1$) and  $b_0=1$ ($J_0=2$).  Error bars  are just the standard error from the mentioned average. The case $\theta=\pi/2$ is presented in  Appendix~\ref{sec:theta90}.}
 
\label{fig:alpha-vs-theta}
\end{figure*}

We devise a protocol for the step that obeys  $J_t = 1 + b_t$ with the binary variable $b=\{b_t\}$. If $J_t=1 \ \forall t$, we recover the Standard QW where the probability flux only occurs towards adjacent sites. In all the cases, we start from $b_0=0$ --- which corresponds to $J_0=1$ --- then we apply one of the following rules:
\begin{itemize}
\item Periodic: the values of $b_t$ are systematically alternated between $0$ and $1$;

\item Fibonacci: the sequence of values of $b$ is generated by applying the substitution rule $0 \rightarrow 01 $ and $1 \rightarrow 0 $;

\item Thue-Morse: the sequence of values of $b$ is generated by applying the substitution rule $0 \rightarrow 01 $ and $1 \rightarrow 10 $;

\item Rudin-Shapiro: first, we generate a sequence with four letters by means of the substitution rule 
$A \rightarrow AB$, 
$B \rightarrow AC$,
$C \rightarrow DB$ and
$D \rightarrow DC$. Then we set $A=B \rightarrow 0$ and 
$C=D\rightarrow 1$;

\item Random: we first generate a periodic sequence until $t_{\max}$, then we randomize it. This procedure is done to make a fair comparison between such protocols.
\end{itemize}

In Fig.\ref{fig:bin-seq}, we display the first $25$ elements of the sequences we have described here. For further details on the Fibonacci, Thue-Morse, Rudin-Shapiro sequences we point the reader to Refs.~\cite{albuquerque2004polaritons,steurer2007photonic,barber2008aperiodic,dal2012deterministic,vardeny2013optics,bellingeri2017optical,lambropoulos2019tight}.

We compute the power spectral density (PSD) for the discrete sequences 
$z_t=b_t$ as shown in Fig.\ref{fig:acf-psd-lzc}(a-c)
\begin{equation}
\Phi (\omega) =  \left| \frac{1}{ \sqrt{2\pi} } \sum_{n=1}^{n=T} z_n e^{-2\pi i \omega n/T }  \right|^2   \quad \omega =1,...,\omega_{\max}
\end{equation}
where the argument of the modulus is the discrete Fourier transform and $\omega_{\max}=T$;
Normalization is implemented in a way that $\sum_\omega \Phi(\omega)=1$. For a proper symmetric analysis of the PSD, we consider the usual procedure of working with the equivalent sequence obtained from a centralization by its mean. 
Recall that for a perfect white noise the autocorrelation function 
\begin{equation}
R(\tau) \equiv \langle z_{t} z_{t-\tau} \rangle  =  \frac{1}{T-\tau} \sum_{t=\tau}^{T} z_{t} z_{t-\tau},
\end{equation}
shall read $R(\tau)=\delta(\tau)$, which yields a flat spectrum $\Phi (\omega )=1$  since all frequencies have the same contribution. Nevertheless, for finite sequences emerges a noisy behavior. We see that the Rudin-Shapiro sequence is broadly scattered over the spectrum. On the other hand, the Fibonacci sequence displays a multi-peaked behavior. The Thue-Morse has an intermediate behavior between both sequences. Alongside the qualitative analysis of Figs.\ref{fig:bin-seq}-\ref{fig:acf-psd-lzc}(a-c) we assess the structural properties of the aperiodic sequences we use for each jump protocol.
As depicted in Fig.~\ref{fig:acf-psd-lzc}(d), with that quantity we reassure the pattern of peaks, which reveals the aperiodic sequences we use have distinct local properties. The overall behavior of the deterministic RS sequence shows that it resembles a purely random sequence, but with much smaller fluctuations.

Evaluating the fraction of $1$s in the binary sequences, we verify that relative frequency of $1$ is strongly unbalanced for Fibonacci chain (see Fig.~\ref{fig:acf-psd-lzc}(e)). The periodic sequence is well-balanced and the random sequence is tailored to be balanced, but it clearly presents local deviation from the unbiased case. The Thue-Morse sequence has the interesting property of being balanced despite its aperiodicity.

In order to further characterize these sequences from a complexity point of view, we considered the evaluation of the Lempel-Ziv complexity, as shown in Fig.~\ref{fig:acf-psd-lzc}(f). That measure computes the number of nonidentical patterns in a sequence when scanned from $t_0$ to $t_{\max}$~\cite{lempel1976complexity}; to that, we use  Kaspar-Schuster's method~\cite{kaspar1987easily} to compute it, see the Appendix~\ref{sec:ks-method-lzc}. The extreme cases in Fig.~\ref{fig:acf-psd-lzc}(f) are the periodic and random sequences with minimum and maximum complexity, respectively. 
Between such extremes, we see the Fibonacci, Thue-Morse, and Rudin-Shapiro with increasing complexity.  
Although the Rudin-Shapiro sequence has a correlation pattern similar to random sequences, it is clear that its Lempel-Ziv complexity is much smaller. That feature is important to explain discrepancies arising in the scaling behavior of the spreading; in other words, the autocorrelation (or the spectral density) is not enough to fully explain our results. That is related to the fact that the disordered sequence has nonlinear dependencies that are not detected by a single measure (see e.g. Ref.~\cite{queiros2009comparative}). \textcolor{black}{Therefore, the application of different aperiodic sequences with different degrees of non-linearity --- which can be understood as complexity as well --- helps shed light on the role of such features on spreading, 
delocalization and entanglement in QWs.}

Last, and for comparison purposes, we have simulated a classical walk using the same protocol of jumps previously defined. Concretely, we use the symmetric discrete-time map
\begin{equation}
P_{t+1}(x) = \frac{1}{2} P_t(x-J_t) + \frac{1}{2} P_t(x+J_t),
\end{equation}
where it is clear the absence of interference effects since the flux of probability from the positions $\{x\pm J_t\}$ do not modulate one another.

\begin{figure*}[t]
     \centering
\includegraphics[scale=0.47]{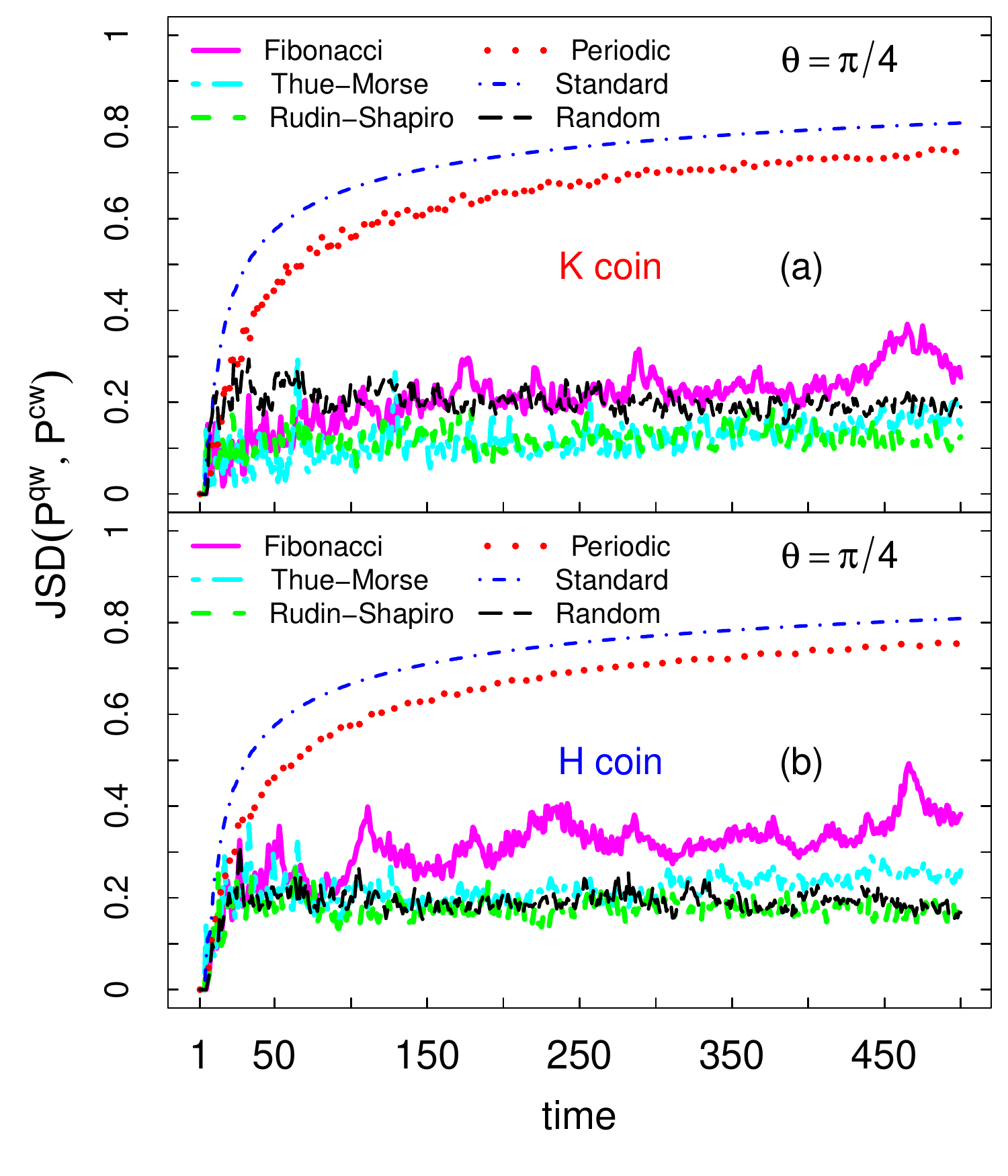}
\includegraphics[scale=0.47]{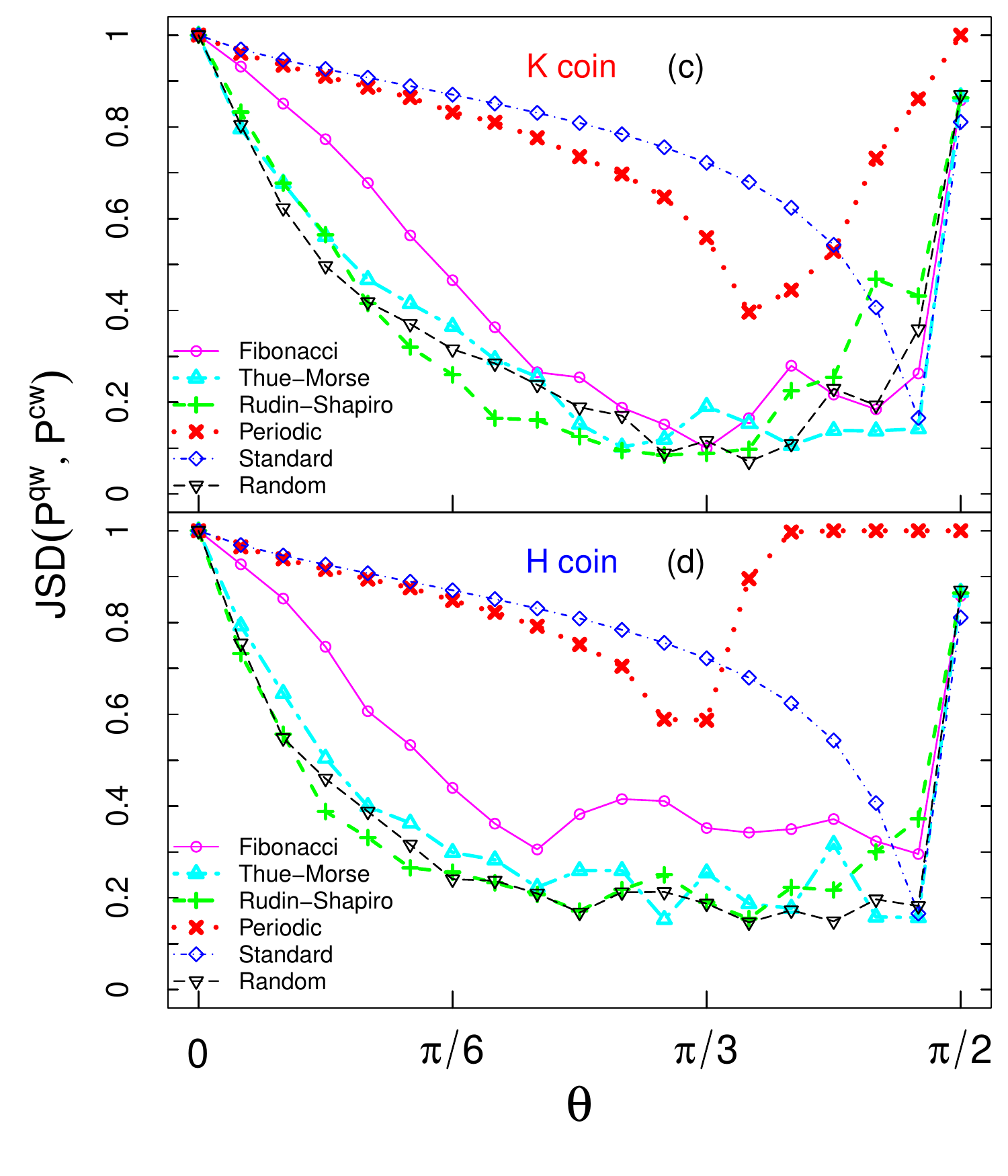}

\includegraphics[scale=0.47]{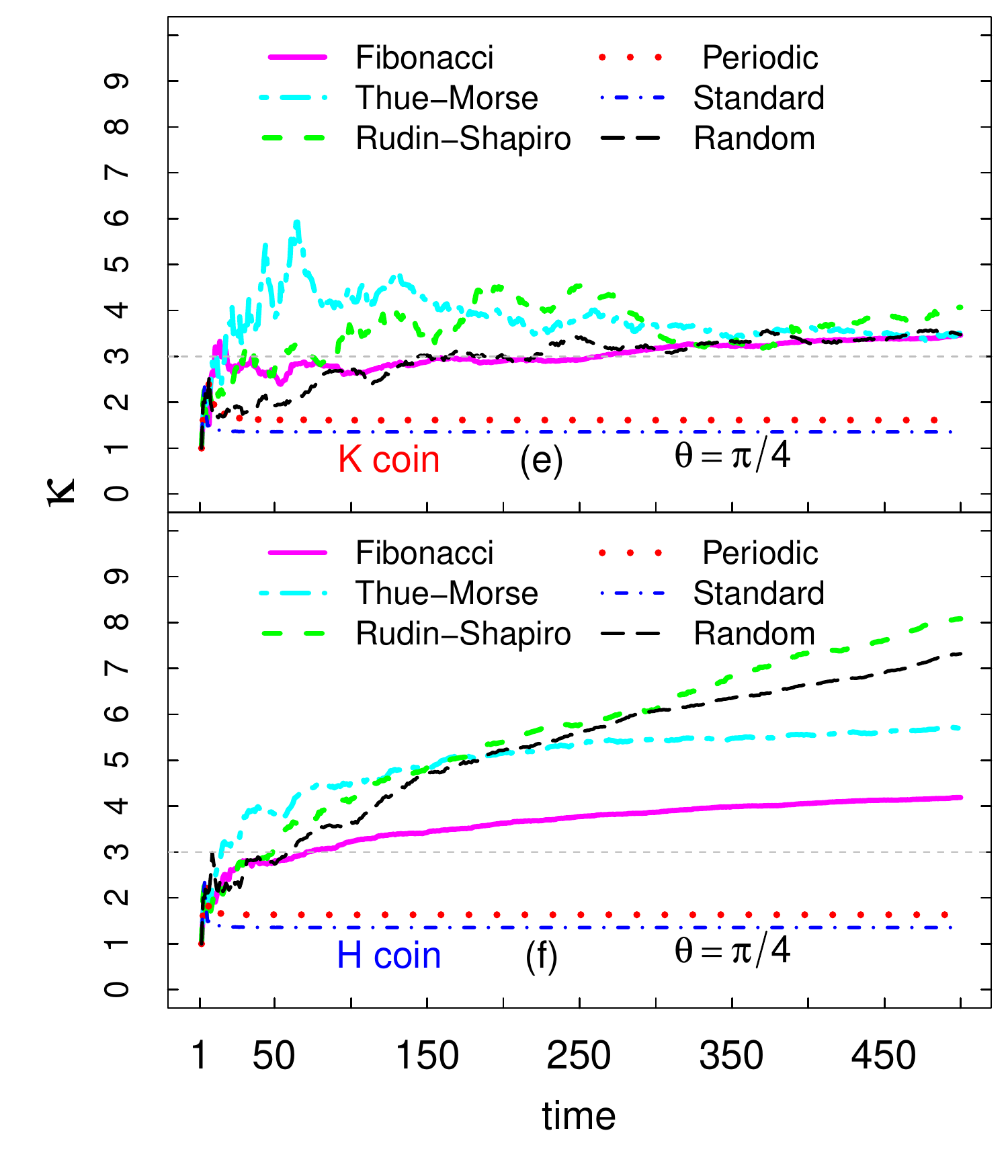}
\includegraphics[scale=0.47]{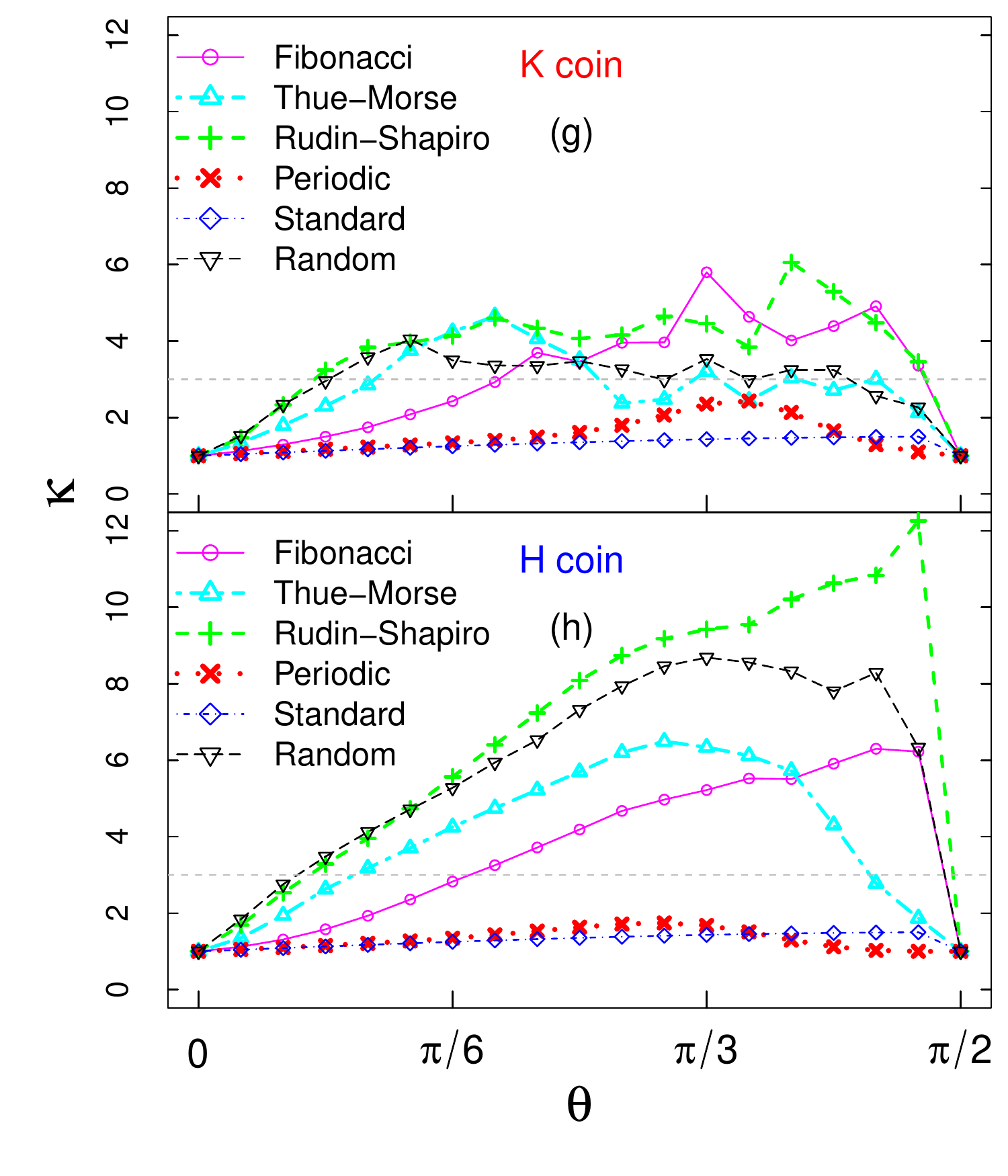}

\caption{(Left) Time series for Jensen-Shannon dissimilarity (JSD) and kurtosis, $\kappa $, for aperiodic/periodic protocols of jumps. (Right) Dependence of such quantities  as a function of $\theta$ at $t=500$.}
\label{fig:jsd-kurt-time-theta}
\end{figure*}

\begin{figure*}[t]
     \centering
\includegraphics[scale=0.47]{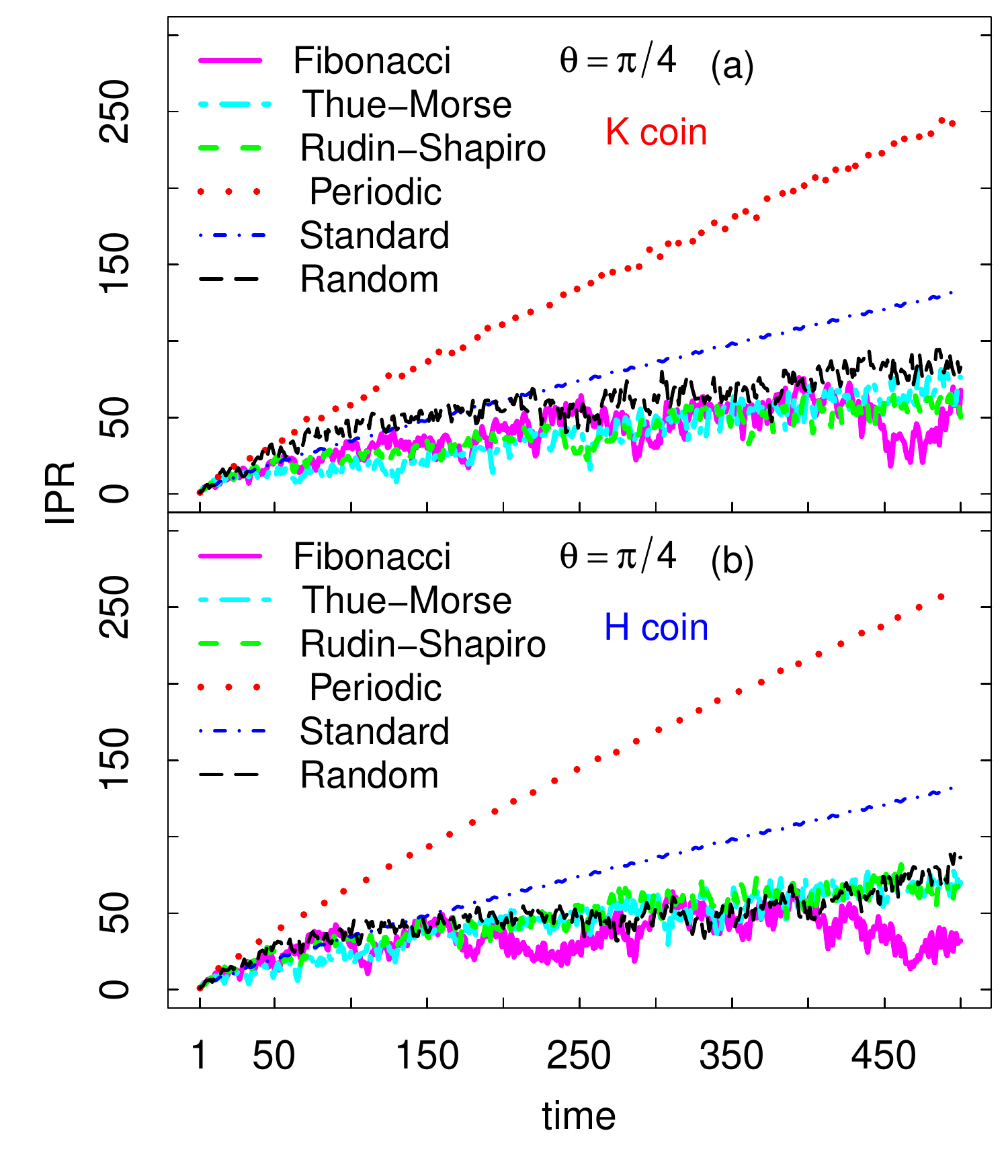}
\includegraphics[scale=0.47]{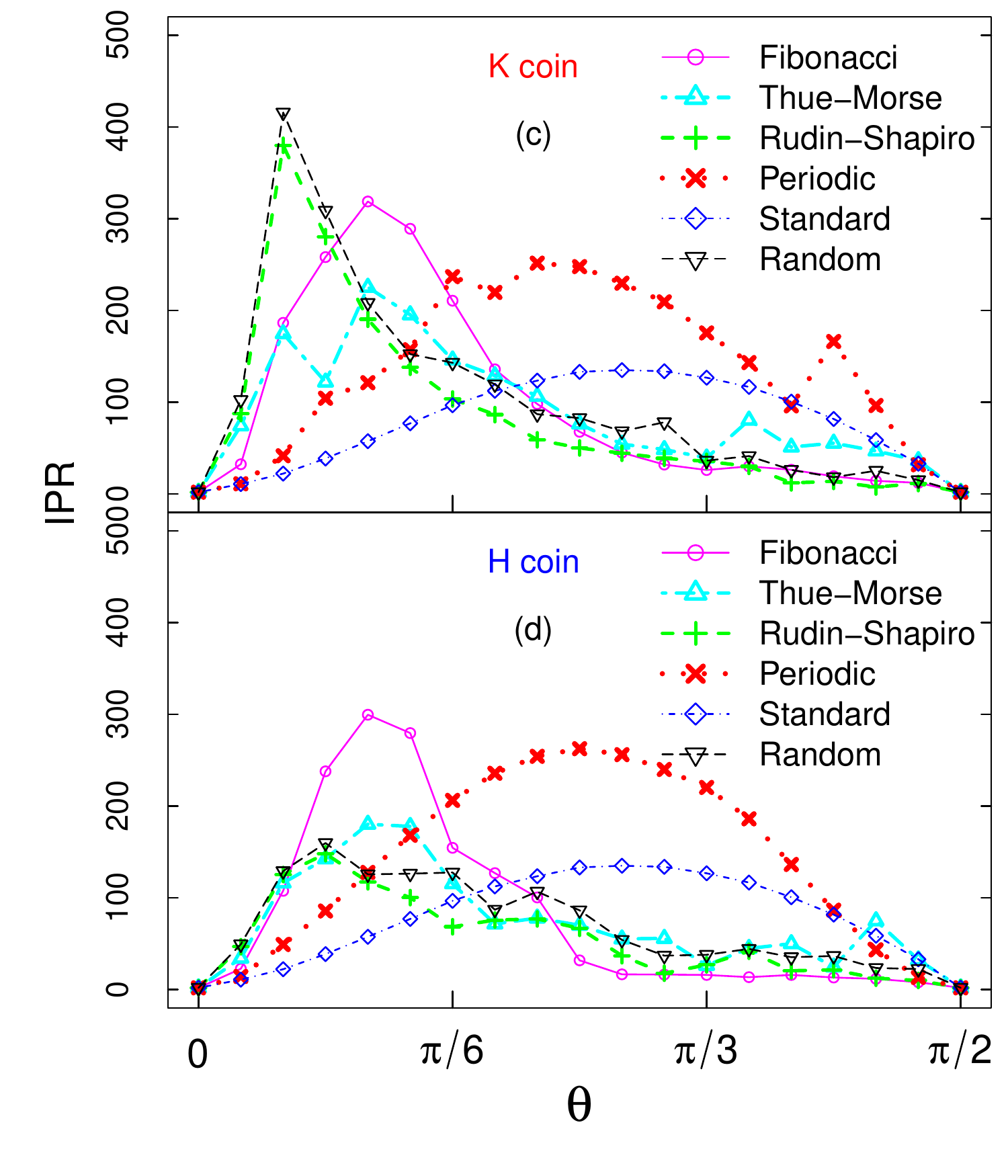}

\includegraphics[scale=0.47]{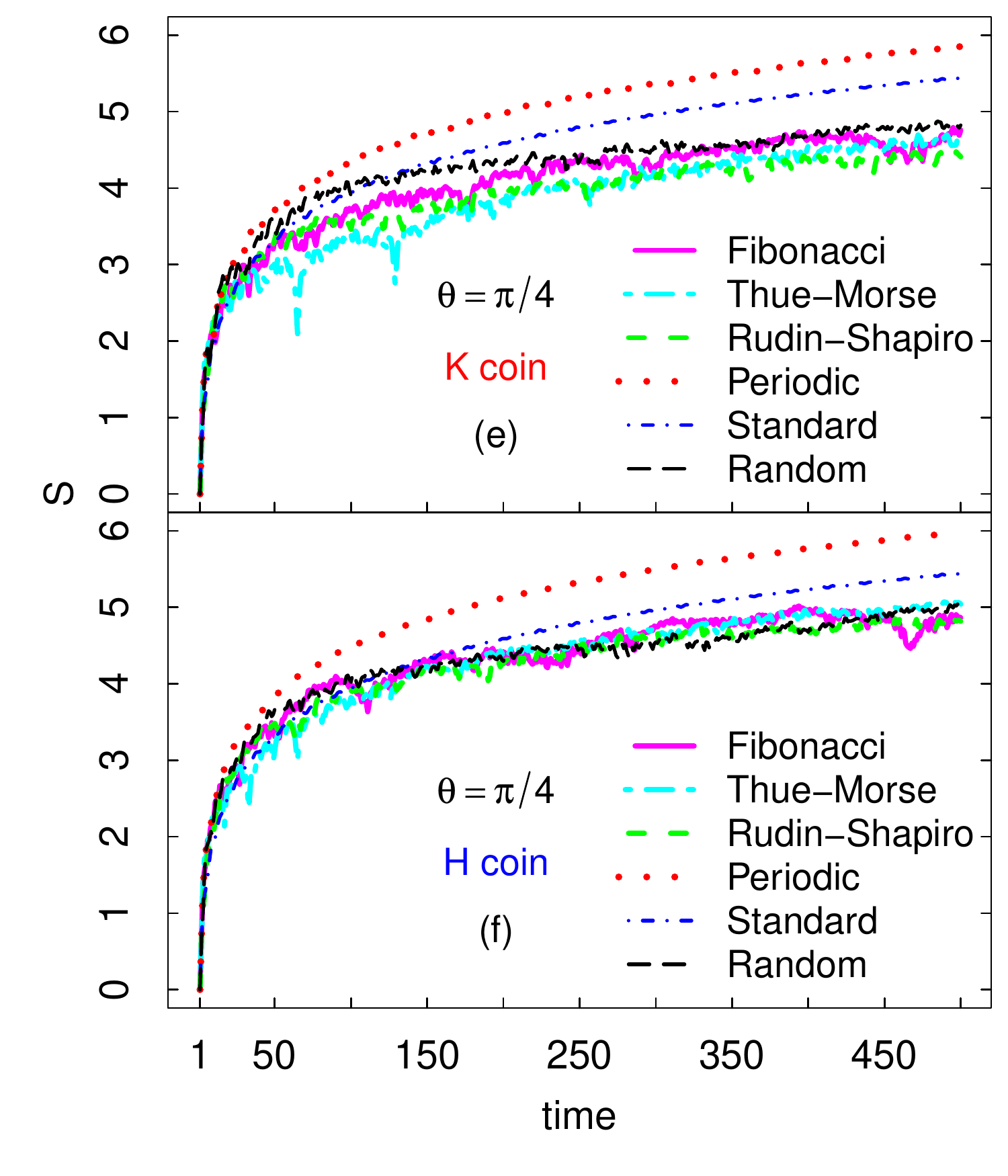}
\includegraphics[scale=0.47]{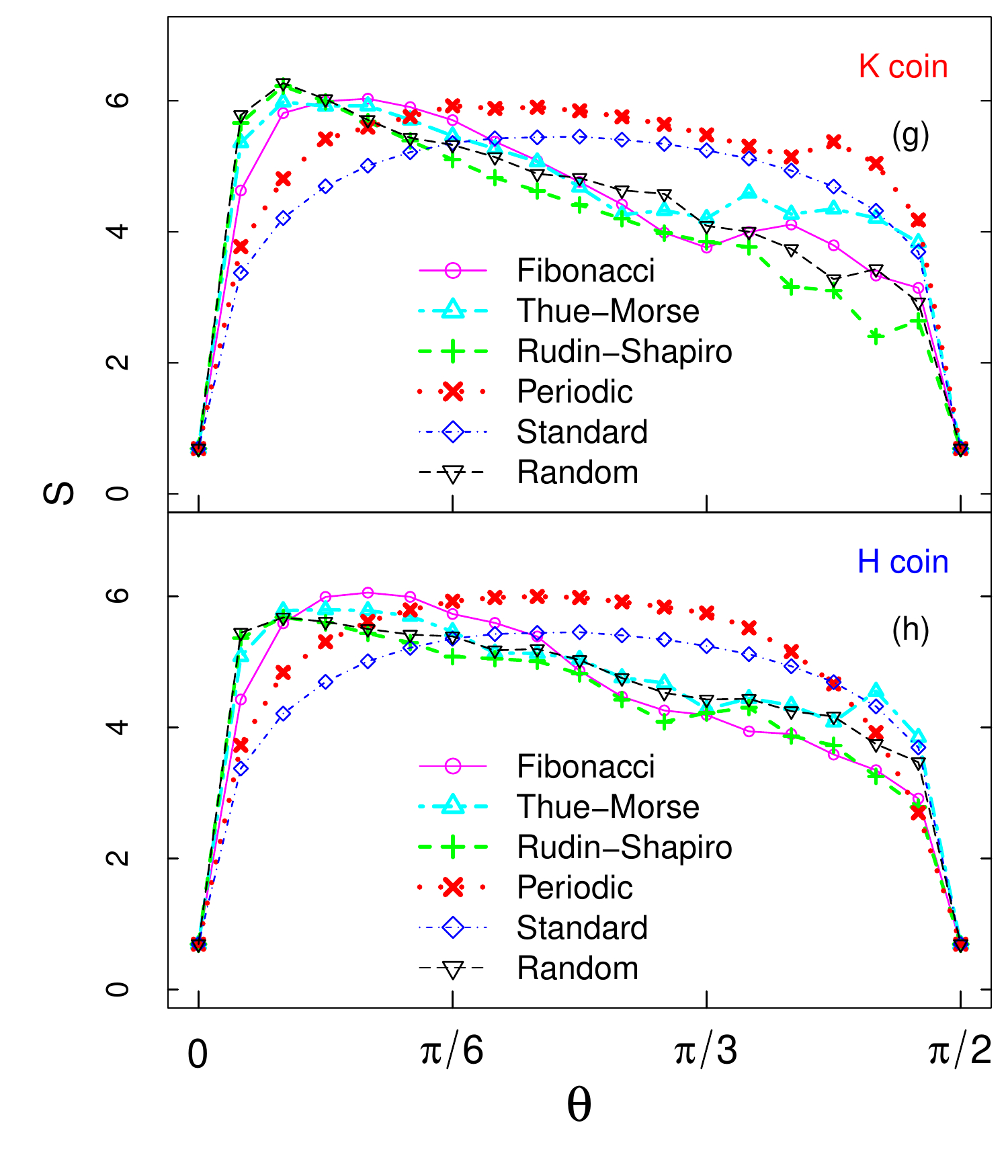}

\caption{(Left) Time series for the  $S$ and IPR for aperiodic/periodic protocols of jumps. (Right) Dependence of such quantities with $\theta$ for $t=500$.}

\label{fig:shan-ipr-time}
\end{figure*}

\begin{figure*}[t]
  \centering
\includegraphics[scale=0.47]{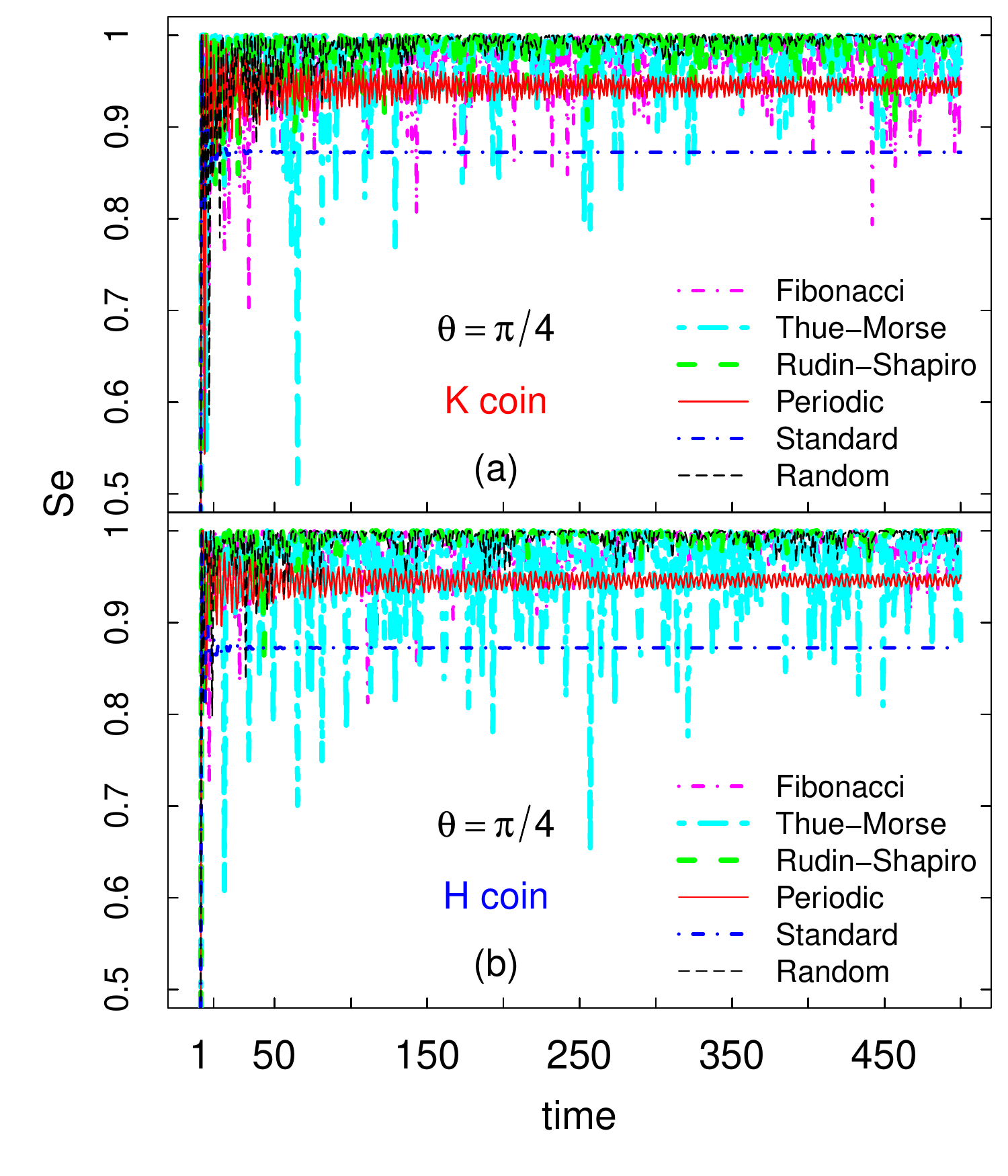}
\includegraphics[scale=0.47]{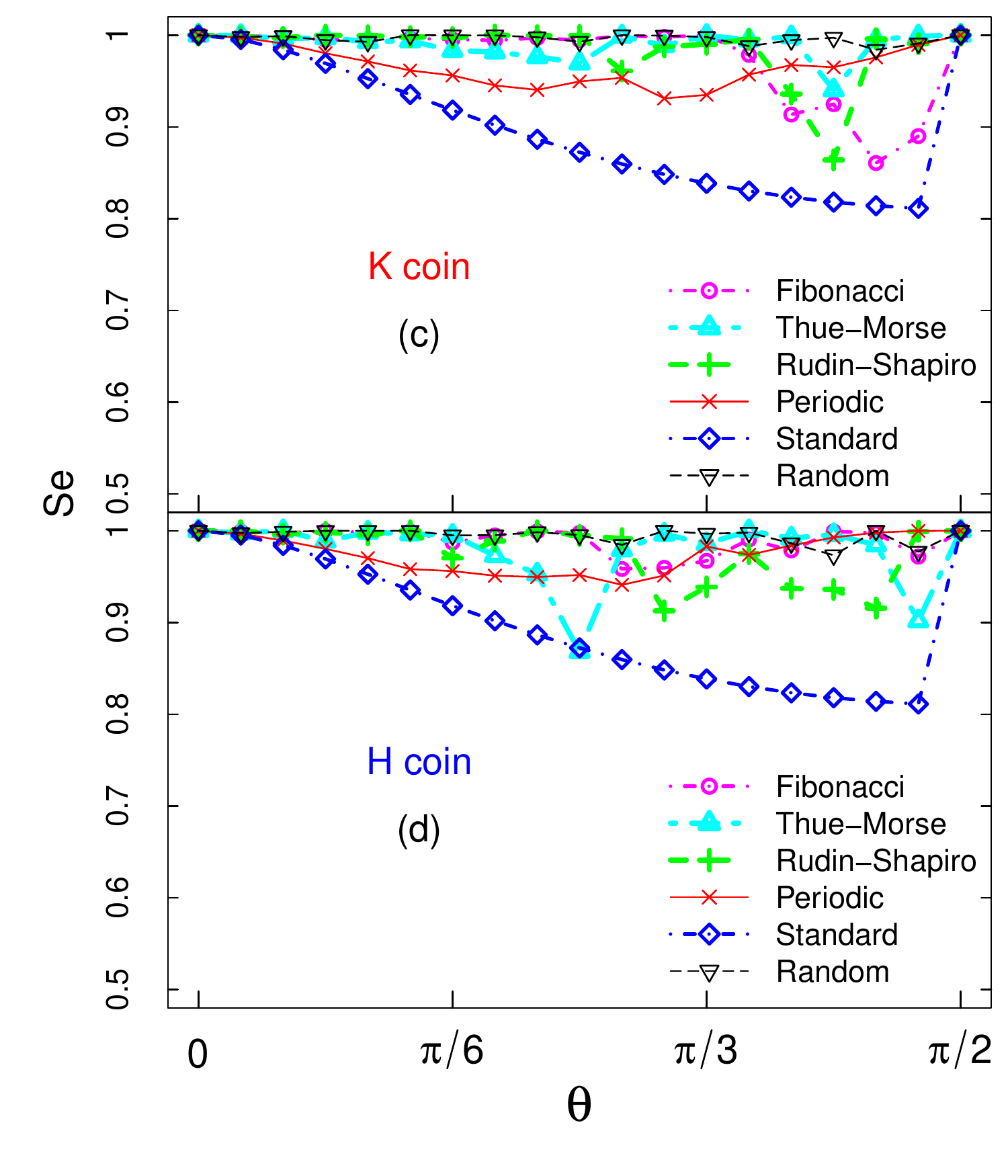}

\caption{Behavior of the von Neumann entanglement entropy $S_e$ in time for $\theta=\pi/4$ (left) and with $\theta$ for $t=500$ (right) for aperiodic protocols. As comparison, we show the time evolution of $S_e$ for the standard case as well as for periodic jumps.}
\label{fig:se-time-theta}
\end{figure*}

\begin{figure*}[t]

      \centering

\includegraphics[scale=0.86]{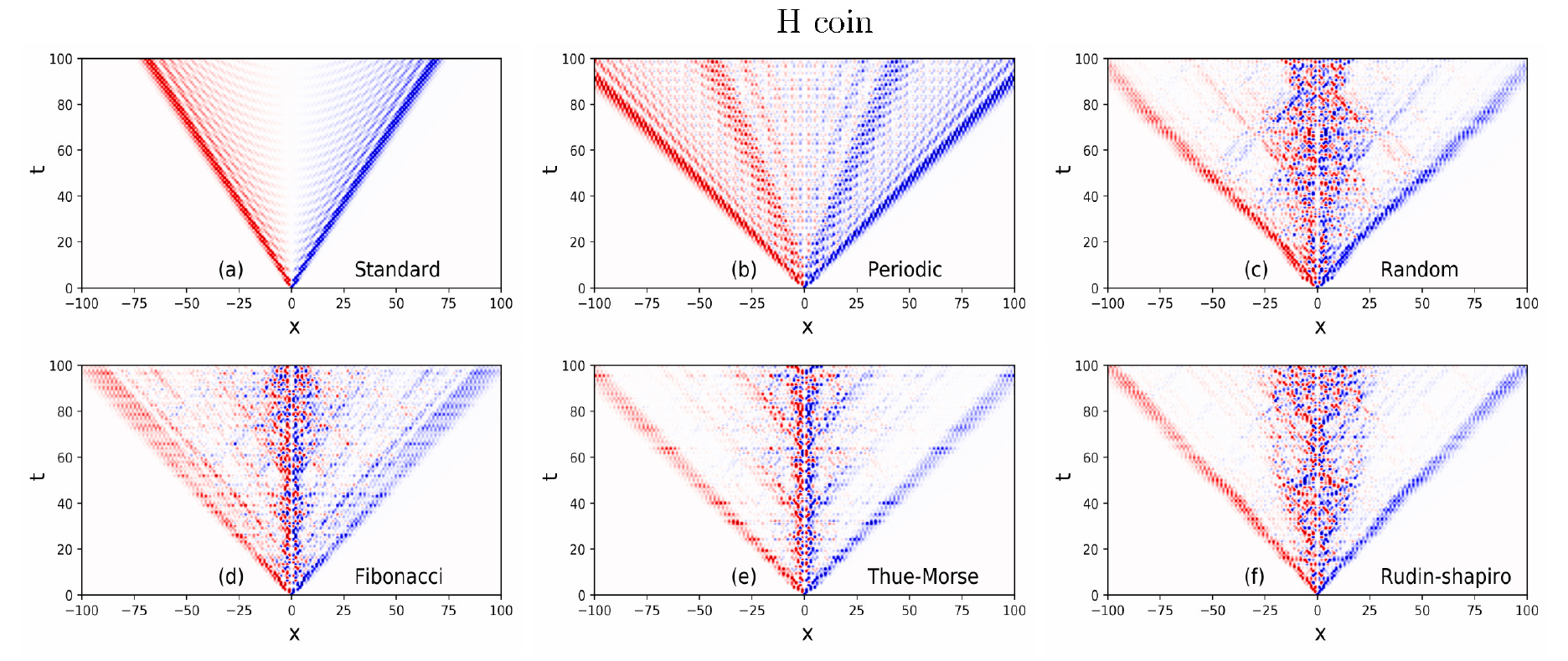}
\includegraphics[scale=0.86]{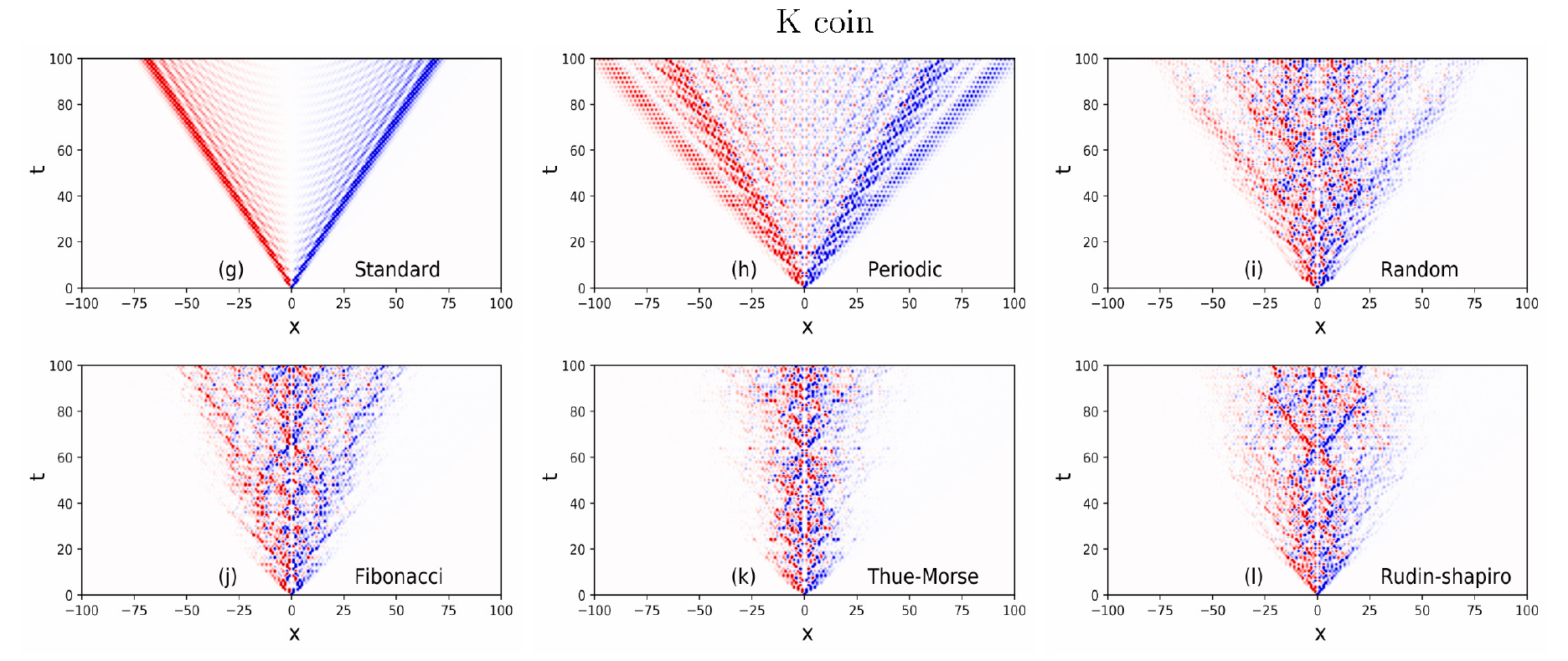}

\caption{Space-time evolution of the normalized asymmetry measure $A_t(x)/|A_t|^{\max}$ for  $\theta=\pi/4$ with the H coin (a-f) and K coin (g-l). Inside each panel the red \textcolor{black}{ (left)} profile highlights the predominance of $|\psi_{t}^{D}(x)|$. Whereas the blue (\textcolor{black}{ right}) profile indicates the prevalence of $|\psi_{t}^{U}(x)|$. In both cases, the darkness denotes the intensity of $A_t(x)/|A_t|^{\max}$. }
\label{fig:mxt45_hk}

\end{figure*}

\section{\label{sec:results}Results and discussion}

In this section, we characterize the global and local properties concerning the evolution of the QW wavefunction. \textcolor{black}{The combination of this analysis with the description of the non-linear correlation features of each aperiodic sequence in Sec.~\ref{sec:model} put us in a position to give an account over the role played by such properties on the characteristics of QWs under those rules.} To accomplish that, we first compute the space-time probability $P_t(x)$ of the corresponding wavepacket
\begin{equation}
P_t(x) = |\psi_{t}^{D}(x)|^2 + |\psi_{t}^{U}(x)|^2 .
\end{equation}

With that result in hand, we compute the $n$-th order statistical moments
\begin{equation}
m_n(t) = \overline{x^n}_t =\sum_x x^nP_t(x) .
\end{equation}

Pivotal for the characterization of each type of quantum walk is the case $n=2$ since as $m_1 = 0 \,\, \forall _t$, $m_2 = \sigma^2 \equiv m_2 - m_1^2 $, which is an effective measure of the wavepacket spreading in time. Typically --- and under the Markov law ---,  physical processes behave asymptotically as $\sigma ^2 (t) \sim t^\alpha \,\, (t\gg 1) $, where the diffusion exponent $\alpha $ is utilized to classify the spreading/diffusion taking place. As we have learnt from the computation of $\sigma ^2 (t)$ plotted in Fig.\ref{fig:x2-qw-crw-log}(a-c), aperiodic jumps play a dual role; on the one hand, they help enhance the spreading in the classical walk, but they induce a counterintuitive inhibition of dispersion for the quantum counterpart as a result of the enhanced interference pattern. On the other hand, we see that the classical spreading keeps the linearity relation $m_2(t) \sim t$. This picture is different from the quantum walks we have analyzed, where the growing speed of $m_2(t)$ seems to change with the kind of aperiodicity. That hint is corroborated with further analysis in 
Fig.\ref{fig:x2-qw-crw-log}(d-f).

In Fig.~\ref{fig:alpha-vs-theta}, we unveil the role of the type of aperiodicity in the scaling behavior of the wavepacket spreading using $t_{\max}=2 \times 10^5$ to estimate the value of $\alpha$.  From that analysis we understand that: (i) the ballistic dynamics --- $\alpha=2$ --- is preserved for periodic jumps;  (ii)  the superdiffusive spreading --- $1<\alpha <2$ --- is present for all deterministic aperiodic sequences in all scenarios with $\theta \neq 0$;  (iii) when $\theta \neq 0$, there is a difference in the value of $\alpha$ depending on the type of coin operator (H or K);  (iv) even though the coin operator is disorder-free ($\theta$ is constant in space-time), we observe the exponent changes nonmonotonically with $\theta$; (v) for processes where the Hadamard coin was applied, $\alpha $ exhibits less variability than in K coin systems; (vi) for the Kempe coin, the results for the Rudin-Shapiro case display clear-cut differences from a purely random setting.

As expected, we have found clear differences in relation to the classical walk model. That comparison is presented in Fig.~\ref{fig:alpha-vs-theta}. Immediately, we see the classical walk instance is not affected by the sort of aperiodic jump protocol we select; nonetheless the quantum approach is slightly sensitive for the Fibonacci and Thue-Morse and strongly sensitive for Rudin-Shapiro implementations, while the CW remains robustly diffusive under aperiodic jumps, we note that the scaling behavior of QW is sensitive to the type of aperiodicity. 
For random jumps with $\widehat C_{H}(\theta=\pi/4)$ we   recover the results shown in~
\cite{sen2019unusual,sen2019scaling,mukhopadhyay2019persistent,das2019inhibition,pires2019multiple}. From that perspective, our random setting generalizes those results for the full range of $\theta \in \{0,\pi/2\}$ with $\widehat C_{H,K}$.

\textcolor{black}{Fig.~\ref{fig:alpha-vs-theta} also shows that the overall values of 
$\alpha$ vs $\theta$ 
decreases as the amount of non-identical patterns increases (LZC increases). Notably, when the LZC is minimum the ballistic spreading ($\alpha=2$) is achieved for both the constant and periodic sequence, which is a surprising result given that the periodic jumps lead the wavepacket to achieve much more distant positions.
These results highlight the important role that the  LZC plays in the scaling behavior of the transport properties. Given the 
relevance of the aperiodic sequences in science and technology \cite{steurer2007photonic,barber2008aperiodic,dal2012deterministic,vardeny2013optics,bellingeri2017optical} and that the LZC is a simple measure to define and compute (see the Appendix~\ref{sec:ks-method-lzc}), we emphasize that the LZC enters as a new control feature for the engineering and manipulating of wavepackets.}

 Still in Fig.~\ref{fig:alpha-vs-theta}, we see another worthwhile result: the implementation with $\theta=0$ is robustly ballistic regardless of the type of temporal disorder in step lengths. Why are aperiodicity-induced effects suppressed for $\theta=0$? The answer to that question is traced back to the mathematical structure of the coin and step operators. When $\theta=0$ the diagonal terms $c_{12}=c_{21}=0$ of the operators $\widehat C_{H,K}$ zero out. The H coin operator becomes the Z-Pauli matrix $\widehat C_{H}= \widehat \sigma_z$, whereas the K coin becomes the identity matrix $\widehat C_{K}=\mathcal{I}$. As diagonal operators now, $\widehat C_{K}$ and $\widehat C_{H}$ furthers the pure propagation of each spin-component in its corresponding direction without interfering with one another. Heeding these features, it becomes clear that under decoupled conditions the ballistic spreading remains safeguarded from the disorder in the step operator. In Appendix~\ref{sec:theta90}, we add a further discussion for the case $\theta=\pi/2$ as well.

 \textcolor{black}{The scaling analysis presented above allows determining the exponent $\alpha$ that is a global measure of the wavepacket transport. Henceforward, we will employ a toolkit from statistics and information theory with the motivation of detecting local distributional fingerprints in the QW dynamics caused by aperiodicity in the hoppings.}

Besides the natural difference in $m_2(t)$ for the quantum and classical walks, we  compute the discrepancy between the  distributions arising from QW and CW,  $P^{{\rm qw}}_{t}(x)$ and $P^{{\rm cw}}_{t}(x)$, by employing tools from the information theory, namely the Jensen-Shannon dissimilarity~\cite{lin1991divergence},
\begin{equation}
JSD_t(P^{{\rm qw}},P^{{\rm cw}}) \equiv 
\frac{KLD_t \left( P^{{\rm qw}} | 
M\right)
+
KLD_t \left( P^{{\rm cw}} |
M \right) 
}{2}  
\end{equation}
where $M(x)$ is the mean distribution 
\begin{equation}
M(x) = \frac{P^{{\rm qw}}(x) + P^{{\rm cw}}(x)}{2}
\end{equation}
and the function $KLD$ is the Kullback-Leibler Dissimilarity,
\begin{equation}
KLD_t(R|W) \equiv \sum_{x} R_t(x) \log_{2} \frac{R_t(x)}{W_t(x)}.
\end{equation}
Among the set of its properties~\cite{briet2009properties}, we emphasize the fact that $JSD_t(P^{{\rm qw}},P^{{\rm cw}})$ has the advantage of being both upper and lower bounded, $0\leq JSD_t(P^{{\rm qw}},P^{{\rm cw}}) \leq 1$ as well as symmetric.
Notwithstanding the recent assertion the Kullback-Leibler measure is very helpful in providing a better understanding of the outcomes arising from their new time-dependent protocol for the coin operator~\cite{panahiyan2018controlling}, we deem symmetric measures like the JSD more reliable.

In Fig.~\ref{fig:jsd-kurt-time-theta}(a-d), we see to what extent $P^{{\rm qw}}_{t}(x)$ and $P^{{\rm cw}}_{t}(x)$ are different due to interference effects.
Specifically, $JSD(P^{{\rm qw}},P^{{\rm cw}})=0$ at $t=0$ since both distributions are equal $P_t^{{\rm qw}}(x)=P_t^{{\rm cw}}(x)= \delta _{x,0}$. Such maximum overlap (minimum dissimilarity) is persistent in the second step because during the initial stage there are not enough quantum states to interfere with one another. That scenery abruptly changes in the subsequent time steps in which emerges an interference-induced breaking in the full similarity $P_t^{{\rm qw}}(x)=P_t^{{\rm cw}}(x)$. That spatial dissimilarity increases quickly in the short-run, but subsequently, its rate peters out. In the right panel, we see a nonmonotonic dependence of $JSD(P^{{\rm qw}},P^{{\rm cw}})$ with $\theta$ for all protocols. The overall behavior of Jensen-Shannon dissimilarity with the type of aperiodicity shows that there is a larger site-to-site overlap (smaller dissimilarity) between $P^{{\rm qw}}_{t}(x)$ and $P^{{\rm cw}}_{t}(x)$ as the complexity of the jump sequence increases.

Further insights on the distributions $P_t(x)$ are obtained  from the behavior of the tails of  $P_t(x)$ on the chain. To that, we employ the kurtosis
\begin{equation}
\kappa \equiv \frac{ m_4(t)}{m_2^2(t) }. 
\end{equation}
In Fig.\ref{fig:jsd-kurt-time-theta}(e-f), 
we see that $\kappa$ exhibits an increasing pattern over time evincing that the core/bulk relationship is changeable. Specifically, this corresponds to a decrease in the relevance of the tails of $P_t(x)$ as the wavepacket spreads on the lattice over time. 
That property comes to happen because these jumps induce two effects: (i)the increase of the range of $x$ satisfying $P_t(x)>0$
; (ii) centralization of  $P_t(x)$. Both effects under the constraint $\sum_x P_t(x)=1$ stimulates the penalization of the importance of tails.
In Fig.\ref{fig:jsd-kurt-time-theta}(g-h), the overall behavior of $\kappa$ vs $\theta$ shows a highly irregular behavior with $\theta$, that is more
pronounced for the K coin. 
For the H coin, it is possible to observe that the weakening in the tails of $P_t(x)$ becomes more pronounced as the complexity of the sequence soars. 
The irregularities in the behavior $\kappa$ are fingerprints of the absence of regularity in the aperiodic sequences.
Such irregularity becomes well visible for $\kappa$ because of its 
quartic polynomial behavior
that contributes to a high sensibility to this measure.
The plots over the evolution of the kurtosis in Fig.~\ref{fig:jsd-kurt-time-theta}(e-h) point to the overall increase of the relevance of the bulk of $P_t(x)$ at expense of the waning of the tails  that is mostly compatible with the overall increase of the overlap between $P^{{\rm qw}}$ and $P^{{\rm cw}}$, which --- in turn --- is stressed by the decrease in $JSD$, as shown in Fig.~\ref{fig:jsd-kurt-time-theta}(a-d). Both features provide complementary information about the slowing down of the QW observed in 
Figs.~\ref{fig:x2-qw-crw-log}-\ref{fig:alpha-vs-theta}.

 We now focus on quantifying the amount of spatial participation of each state in the total wavepacket. 
For this task, two common quantities  can be  employed, namely the
Shannon entropy (S)~\cite{lavivcka2011quantum,panahiyan2018controlling,ahmad2019one,bhandari2019light} and the Inverse participation ration (IPR)~\cite{ghosh2014simulating,yalccinkaya2015two,zeng2017discrete,derevyanko2018anderson,buarque2019aperiodic} of the probability profile. Explicitly, 
\begin{equation}
S \equiv  - \sum_x   P_t(x)  \log  P_t(x),
\end{equation}
and
\begin{equation}
IPR \equiv  \left( \sum_x (P_t(x))^2 \right)^{-1},
\end{equation}
respectively. Those measures allow detecting different spatial features of the wavepacket delocalization. 
\textcolor{black}{Although the Shannon entropy, $S$, is notoriously a classical quantity, in our case it can be associated with the delocalization of $P_t(x)$ over the chain within the context of QWs in the sense it has 2 well-defined extremes:
(i) fully localized states $ \rightarrow P_t(x)=\delta_{x,0} \rightarrow S=0$;
(ii) fully delocalized states$ \rightarrow P_t(x)=1/N \rightarrow S=\log N$ where $N$ is the maximum possible number of sites in which $P_t(x)$ can be distributed. In other words, we read $S$ a distributional measure that gives us a complementary insight into how much contribution each state provides to the full $P_t(x)$ and thus the impact of linear and non-linear correlations in this  feature. The same sort of knowledge is given by the Inverse Participation Ratio with $IPR = 1$ indicating fully localization whereas $IPR = N$ corresponds to complete delocalization.}

In Fig.~\ref{fig:shan-ipr-time},  we perceive  that $IPR$ is more wobbly than the entropy because of its quadratic behavior that leads to high sensibility to spatiotemporal variations in $P_t(x)$. Such feature is smoothed in the figures provided by the calculations of the Shannon entropy, which --- because of its logarithmic dependence --- assigns little weight to the sites with $P_t(x)<<1$. 
For both measures, we see a highly non-trivial dependence on $\theta$ which is one of the outcomes of the irregular presence of aperiodic jumps. That irregularity in the jumps arises from the absence of regularity of the corresponding aperiodic sequences.
For the standard QW setting, we recover the smooth curve for $S$, as previously obtained~\cite{chandrashekar2008optimizing}. The nonmonotonic shape for $S$ and $IPR$  can be explained by the modulation of the competition between two mechanisms: (i) as $\theta \rightarrow 0 $, the spreading of the wavepacket is enhanced, which permits new sites significantly off the origin to participate in $P_t(x)$; (ii) as  $\theta $ approaches the unbiased coin case $\theta = \pi/4$, spatial splitting of states becomes more balanced between the spin components $\ket{\downarrow}$ and $\ket{\uparrow}$, thus allowing old sites near the origin to keep a non-negligible contribution to $P_t(x)$.

\textcolor{black}{Focusing on the role of the complexity, we see in Fig.\ref{fig:shan-ipr-time}(c-d,g-h)  that both the entropy and IPR become left-skewed for all types of sequences with non-trivial patterns. That indicates the delocalization of the wavepacket is increased towards $\theta<\pi/4$, which are angles of $\widehat C_{H,K}$ that favor the components related to propagation ($c_{11}, c_{22}$).
For $\theta=0$, the spreading is ballistic but with only two sites participating in the full wavepacket; now our results show that by choosing protocol of jumps with nontrivial complexity it is possible to attain a superdiffusive regime with the extra possibility 
for tuning $\theta$ in order to increase propagation without too much loss in the spatial participation of the local spinors in the total wavepacket. That finding offers a new possibility for tuning both the delocalization and propagation of QWs.}

 \begin{center}
\begin{table*}[t]
     \centering
     \begin{tabular}{{ |p{6cm}|p{8cm}|  }}
     \hline
    \hspace{0.9cm}  Time evolution & \hspace{2.2cm} Scaling behavior \\\hline
    CW: enhancement of spreading & CW: $\alpha$ invariant regardless  the type of aperiodicity \\\hline
    QW: inhibition  of spreading & QW: $\alpha$ depends on the type of aperiodicity
     \\
     \hline
     \end{tabular}
     \caption{\textcolor{black}{Comparison between CWs (Classical Walks) and QWs (Quantum Walks) under the effects of aperiodic jumps.}}
     \label{tab:comarison_ca_qw}
 \end{table*}
 \end{center}
 
Qubit-lattice entanglement is another important feature in the evolution of a quantum walk. To quantify this property, we compute the von Neumann entropy
\begin{equation}
S_e  \equiv - \mathrm{Tr} \left[ \rho^c\log\rho^c \right].
\label{S_e}
\end{equation}
To that, we must have the full density matrix $\rho = | \Psi \rangle \langle  \Psi | $ of the QW system whence we obtain the reduced density matrix of the quantum walker
\begin{equation}
\rho^c \equiv {\rm Tr}_x(\rho).
\label{densitymatrix}
\end{equation}
where ${\rm Tr}_x$ stands for the trace over the position base.
\textcolor{black}{Since $\rho^c $ involves tracing out the position degree of freedom, then $0\leq S_e\leq 1$ is interpreted as  a quantifier of the  entanglement between the internal degree of freedom of the quantum walker (spin, polarization) and the external degree of freedom (chain). }
Explicitly, considering Eq.~(\ref{wavefunction}) and following the same steps as in~\cite{abal2006quantum,vieira2013dynamically,zeng2017discrete} the reduced density matrix reads
\begin{align}
\rho^c
= &
\begin{bmatrix}
G_a & G_{ab} \\
G^{*}_{ab} & G_b
\end{bmatrix}
= \\ = &
\sum_x
\begin{bmatrix}
|\psi_{t}^{D} (x)|^2                                 & \psi_{t}^{D}(x)  \left( \psi_{t}^{U}(x)  \right)^{*} \\
\psi_{t}^{U}(x)  \left( \psi_{t}^{D}(x)  \right)^{*} & |\psi_{t}^{U} (x)|^2 
\end{bmatrix}
\end{align}
wherefrom we compute the eigenvalues $\lambda^{\pm}$ of $\rho^c$,
\begin{equation}
 \lambda^{\pm} = \frac{1}{2} \pm \frac{1}{2}\sqrt{1-4G_aG_b + 4|G_{ab}|^2 },
\end{equation}
which finally yields the entanglement entropy,
\begin{equation}
 S_{\rm e}  = -\lambda^{-} \log_2 \lambda^{-} - \lambda^{+} \log_2 \lambda^{+}
\end{equation}

In Fig.~\ref{fig:se-time-theta}, we present how much entanglement is generated by the sequential application of the coin and translation operator with jumps. Remind that in all cases the QW starts from a separable state $S_{\rm e}=0$. 
For the disorder-free setting, $S_{\rm e} \rightarrow 0.872\ldots $ in agreement with~\cite{carneiro2005entanglement,abal2006quantum}.
All the disordered settings, deterministic or random, leads to a jump-induced enhancement of the spin-space entanglement. 
Notwithstanding, aperiodicity makes the entanglement more susceptible to fluctuations. These features are robustly present in the right panel where we show $S_{\rm e} $ vs $\theta$. 
\textcolor{black}{Taking a closer look at $S_{\rm e} $ vs $\theta$ for the settings with the Rudin-Shapiro and random we observe that 
$S_{\rm e}^{rs}$ tends to be smaller than 
$S_{\rm e}^{rand}$. At first, that is intriguing given that both RS and random series have no linear self-dependence (Pearson's correlation is null).  On the one hand, that shows the hidden nonlinear dependencies in the sequence of jumps -- which are not detected by the ACF and PSD~\cite{queiros2009comparative} -- become visible in the entanglement measure.  On the other hand, that same result shows the remarkable role of randomness plays in the nature of generation of entanglement.}

Aiming at better grasping the underlying mechanism behind all these results, we have evaluated the space-time asymmetry~\cite{souza2013coin} between the spin components
\begin{equation}
A_t =  |\psi_{t}^{U}(x)|^2 - |\psi_{t}^{D}(x)|^2 ,
\label{eq:assymetry}
\end{equation}
which assesses the flux of probability through the lattice. Taking into account that one assume that the patterns formed in space-time are more relevant than the magnitude of $A_t$ in itself,  we have plotted in Fig.~\ref{fig:mxt45_hk} the evolution of $A_t(x)/|A_t|^{\max}$ where $|A_t|^{\max}  = \max_{x} A_t $ is the maximum over the chain for each time step $t$; therein, it is possible to perceive at every time step each trajectory is constantly branching due to the transformation of each state into a superposition of other states. 
In the panels (b-f) of Fig.~\ref{fig:mxt45_hk}, we observe persistent secondary peaks near the borders as well, a property that is mildly reminiscent of the ballistic spreading, which is less pronounced for the K coin as shown in Fig.~\ref{fig:mxt45_hk}(h-l). 
The peaks close to the edges are weaker than those for the H coin or are absent at all. This arises --- as shown in such  quantum carpets --- as a result of the enhanced interference pattern between the components  $|\psi_{t}^{D}(x)|$ and  $|\psi_{t}^{U}(x)|$. The presence or not of such off-center peaks is the main origin of the differences in the scaling exponents: $\alpha_{H}>\alpha_{K}$ in general.

 \textcolor{black}{At this point it is worthwhile to discuss the role of the fraction of $1$s in the binary sequences. To this task, consider the periodic and TM sequences that are  well-balanced (Fig.~\ref{fig:acf-psd-lzc}(e)) but exhibit different complexity (Fig.~\ref{fig:acf-psd-lzc}(f)). Despite both sequences having the same number of jumps, in all the measures we considered, we did not find the same dynamical features exhibited by the cases we have studied. That is, the absence of periodicity  in the TM protocol leads to noticeable dynamical differences with respect to the 50/50 periodic sequences. That result reveals that sequence complexity plays a much more important role than the fraction of $1$s in the sequences of jumps. This is an important knowledge to have before investing time and resources in the design of new experimental setups.}

\section{\label{sec:remarks}Concluding remarks}

While for aperiodic disorder in the coin operator of a quantum walk process there are recent works conveying an augment of entanglement provided by the application of aperiodic protocols~\cite{liu2018entanglement,buarque2019aperiodic}, the approach we have implemented herein base on deterministic  aperiodic disorder in the step operator --- and the corresponding enhancement of entanglement ---  is novel to the literature.

For the random  protocol, there is a series of works reporting entanglement production with either disorder in the coin  operator~\cite{chandrashekar2012disorder,vieira2013dynamically,vieira2014entangling,di2016discrete,chakraborty2017quantum,zeng2017discrete,kumar2018enhanced,orthey2019weak} or in the step operator~\cite{sen2019scaling,mukhopadhyay2019persistent,pires2019multiple}. Recently, it was experimentally 
verified that dynamic disorder in the coin operator can lead to  enhancement of entanglement in photonic quantum walks~\cite{wang2018dynamic}. For disorder in the steps, there are no experimental results so far, but our protocol is a potential candidate thereto since very short-range jumps can be implemented with the optical multi-ports platform theoretically proposed in~\cite{lavivcka2011quantum}. The deterministic character of our setting is another advantage since it avoids a sampling process that is challenging for experimentalists as discussed in~\cite{nguyen2019localized,nguyen2019quantum}.

As QWs can be realized with integrated optical waveguide devices\cite{wang2013physical}, our work offers new perspectives for the  developments of new  photonic architectures\cite{dreisow2008second} with aperiodic second-order coupling.

\textcolor{black}{In terms of quantum  transport, we stress that our protocol provides a novel  mechanism for the emergence of superdiffusive spreading, i.e., a regime belonging to the broad class of anomalous diffusion\cite{oliveira2019anomalous}.
An additional feature of our proposal assuming aperiodic jumps is that by  controlling $\theta$ the setup offers the possibility of tuning the scaling exponent $\alpha$ in a given range of values in the class of superdiffusion. Such results naturally prompt the definition of new aperiodic sequences for allowing a broader adjustment of the scaling exponent $\alpha$.}

For  comparison purposes, we established a classical version of the quantum protocols as well. Concerning the time-dependent quantities, the variance of the QW distribution experiences a slowdown with next-nearest neighbor hopping,  whereas the variance of the CW increases with aperiodic jumps. Concerning the asymptotic behavior (large $t$) of the variance, the QW variants we introduced reveal the aperiodicity-driven sensitivity of the scaling exponent. \textcolor{black}{A summary is shown in Table \ref{tab:comarison_ca_qw}.}

Besides the second statistical moment --- from which we analyzed the spreading features ---,  we also highlight the applicability of a set of tools from statistics and information theory in providing a deeper understanding of the underlying space-time features of the QW probability distribution. Specifically, we employ the  Shannon entropy, IPR, Jensen-Shannon dissimilarity and kurtosis. Such distributional measures allow grasping to what extent the changes in the functional shape of $P^{{\rm qw}}_t(x)$ relate to the dynamical behavior of the QW. \textcolor{black}{Our results show that by making a judicious choice of the aperiodic sequence of jumps and $\theta$ it is possible to induce changes in the core/bulk relationship of $P^{{\rm qw}}_t(x)$ in a way that favors propagation as well as an enhanced  participation of the local spinors in the full wavepacket. This finding reveals that the setup introduced in this manuscript enlarges the range of possibilities for tuning delocalization and propagation of QWs. This discovery is worth of attention given that in certain algorithmic instances is desired to increase delocalization without too much loss in propagation\cite{kendon2003decoherence}.}

Last, these results demonstrate that QWs can distinguish the complexity of the sequence used as disorder, Fig.~\ref{fig:alpha-vs-theta}. This feature is more noticeable for the uncorrelated sequences in the panels Fig.~\ref{fig:alpha-vs-theta}(e-f) where $\alpha_{RS} > \alpha_{\rm random} $. That is to say, QWs help detect the intrinsic nonlinearities in the RS sequence that resembles the power spectrum of random sequences. 
If we recall that classical walks are very helpful in a variety of pattern recognition tasks\cite{xia2019random}; the present results suggest further new applications of quantum walks, namely as a tool for analyzing the nonlinearity of time series. 
Such a task of bridging QWs with pattern detection is one of our subsequent avenues of research. 
Already existing results~\cite{emms2009coined} point out the beneficial intersection of both fields.

\begin{acknowledgments}
We acknowledge Giuseppe Di Molfetta for multiple discussions on quantum walks and the financial support from the Brazilian funding agencies CAPES (MAP) as well as CNPq and FAPERJ (SMDQ). 
\end{acknowledgments}

\appendix
\section{\label{sec:ks-method-lzc}KS method for Lempel-Ziv complexity}

The KS approach~\cite{kaspar1987easily} consists of the proper
use of elementary string manipulation tools: concatenation, insert, delete and search for substrings.

We first define the main binary string of size $n$ as $\overline{M}=s_1s_2s_3,..., s_n$, where we call the minimal substring $s_i$ a character or digit.
We define an empty ancillary string $\overline S$ and an empty ancillary substring $\overline Q$ as well. The LZC is started at $c=1$.
Access the first character of $\overline M$ and include it in the ancillary string: $\overline S=s_1$.
Include the second character of $\overline M$ in the ancillary substring: $\overline Q=s_2$.  
Now we have $\overline S=s_1$ and $\overline Q=s_2$. 
Create $\overline{SQ}$, the concatenation of $\overline S$ and $\overline Q$.
Create $\overline{SQ}\pi$ by removing the final character from $\overline{SQ}$.
That is $\pi$ is defined as the operation that removes the last character of a string or substring.
Ask the question: is $\overline{Q}$ contained in the vocabulary $v(\overline{SQ}\pi)$?
If the answer is positive, $\overline{Q} \in v(\overline{SQ}\pi) $, then append $s_3$ into $\overline{Q}$ that now becomes
$\overline{Q}=s_2s_3$. Repeat the previous steps $\overline{SQ} \rightarrow \overline{SQ}\pi $ and ask the question
$\overline{Q} \in \overline{SQ}\pi$? When a negative answer is given, that is $\overline{Q} \notin v(\overline{SQ}\pi)  $, then append  $\overline{Q}$ into the $\overline S$. Increase the LZC: $c \rightarrow c+1$. 
With the new ancillary substring, $\overline S$ start the operations of
concatenation, delete, search of substrings described above.
If the substring  $\overline{SQ}$ reached the size of the main string $\overline{M}  $ the algorithm ends and we set the last increase in the LZC: $c \rightarrow c+1$. 

Now let us show some examples of such an approach.

Example 1: constant sequence $\overline{M} = 111$:

\begin{itemize}

\item  Set $c=1$. The first character has always to be included  $\rightarrow 1 \cdot $, where the dot $\cdot$ means newly inserted character. 

\item $\overline S=1$, $\overline Q=1$, $\overline{SQ}=11$, $\overline{SQ}\pi=1$,  $\overline{Q} \in v(\overline{SQ}\pi) $
$\rightarrow 1 \cdot 1$
\item $\overline S=1$, $\overline Q=1\textcolor{blue}{1}$, $\overline{SQ}=111$, $\overline{SQ}\pi=11$,  $\overline{Q} \in v(\overline{SQ}\pi) $
$\rightarrow 1 \cdot 1  \cdot$

\item The end of $\overline{M}$ has been reached, so $c \rightarrow c+1$ then $c=2$ is the LZC of a constant sequence, that indeed is the minimum possible value. This means that we only need to insert the first value $k$ and then the whole sequence can be reconstructed by copying such character: $ kkkk\ldots \rightarrow k\cdot kkk\ldots$
\end{itemize}

Example 2: periodic sequence $\overline{M} = 1010$:

\begin{itemize}

\item  Set $c=1$ and access the first character $\rightarrow 1 \cdot $

\item $\overline S=1$, $\overline Q=0$, $\overline{SQ}=10$, $\overline{SQ}\pi=1$,  $\overline{Q} \textcolor{red}{\notin} v(\overline{SQ}\pi) $
$\rightarrow 1 \cdot 0   \cdot $, then $c \rightarrow c+1 \Rightarrow  c=2$

\item $\overline S=1\textcolor{red}{0}$, $\overline Q=1$, $\overline{SQ}=101$, $\overline{SQ}\pi=10$,  $\overline{Q} \in v(\overline{SQ}\pi) $
$\rightarrow 1 \cdot 0 \cdot 1  \cdot$

\item $\overline S=10$, $\overline Q=1\textcolor{blue}{0}$, $\overline{SQ}=1010$, $\overline{SQ}\pi=101$,  $\overline{Q} \in v(\overline{SQ}\pi) $
$\rightarrow 1 \cdot 0 \cdot 1 \cdot 0  \cdot$

\item The end of $\overline{M}$ has been reached, so $c \rightarrow c+1 \Rightarrow c=3$ as shown in Fig.\ref{fig:acf-psd-lzc}(e).

\end{itemize}

Example 3\cite{kaspar1987easily}:  $\overline{M} = 0010$:

\begin{itemize}

\item  Set $c=1$ and start with $\rightarrow 0 \cdot $

\item $\overline S=0$, $\overline Q=0$, $\overline{SQ}=00$, $\overline{SQ}\pi=0$,  $\overline{Q} \in v(\overline{SQ}\pi) $
$\rightarrow 0 \cdot 0  \cdot$ 

\item $\overline S=0$, $\overline Q=0\textcolor{blue}{1}$, $\overline{SQ}=001$, $\overline{SQ}\pi=00$,  $\overline{Q} \textcolor{red}{\notin} v(\overline{SQ}\pi) $
$\rightarrow 0 \cdot 01   \cdot $, then $c \rightarrow c+1 \Rightarrow  c=2$

\item $\overline S=0\textcolor{red}{01}$, $\overline Q=0$, $\overline{SQ}=0010$, $\overline{SQ}\pi=001$,  $\overline{Q} \in v(\overline{SQ}\pi) $
$\rightarrow 0 \cdot 01 \cdot 0  \cdot$

\item The end of $\overline{M}$ has been reached, so $c \rightarrow c+1 \Rightarrow c=3$. This leads to the 
partitioning $\overline{M} = 0010 \Rightarrow 0 \cdot 01 \cdot 0 $ where $c=3$ is the number of partitions separated by dots.

\end{itemize}

Example 4: Fibonacci chain $\overline{M} = 10110101$:

\begin{itemize}
\item  Set $c=1$ and begin with $\rightarrow 1 \cdot $

\item $\overline S=1$, $\overline Q=0$, $\overline{SQ}=10$, $\overline{SQ}\pi=1$,  $\overline{Q} \textcolor{red}{\notin} v(\overline{SQ}\pi) $
$\rightarrow 1 \cdot 0  \cdot $, then $c \rightarrow c+1 \Rightarrow  c=2$

\item $\overline S=1\textcolor{red}{0}$, $\overline Q=1$, $\overline{SQ}=101$, $\overline{SQ}\pi=10$,  $\overline{Q} \in v(\overline{SQ}\pi) $
$\rightarrow 1 \cdot 0 \cdot 1$. 

\item $\overline S=10$, $\overline Q=1\textcolor{blue}{1}$, $\overline{SQ}=1011$, $\overline{SQ}\pi=101$, $\overline{Q} \textcolor{red}{\notin} v(\overline{SQ}\pi) $
$\rightarrow 1 \cdot 0 \cdot 11  \cdot $, then $c \rightarrow c+1 \Rightarrow  c=3$

\item $\overline S=10\textcolor{red}{11}$, $\overline Q=0$, $\overline{SQ}=10110$, $\overline{SQ}\pi=1011$, $\overline{Q} \in v(\overline{SQ}\pi) $
$\rightarrow 1 \cdot 0 \cdot 11 \cdot 0  \cdot $

\item $\overline S=1011$, $\overline Q=0\textcolor{blue}{1}$, $\overline{SQ}=101101$, $\overline{SQ}\pi=10110$, $\overline{Q} \in v(\overline{SQ}\pi) $
$\rightarrow 1 \cdot 0 \cdot 11 \cdot 0 \cdot 1  \cdot $

\item $\overline S=1011$, $\overline Q=01\textcolor{blue}{0}$, $\overline{SQ}=1011010$, $\overline{SQ}\pi=101101$, $\overline{Q} \textcolor{red}{\notin} v(\overline{SQ}\pi) $
$\rightarrow 1 \cdot 0 \cdot 11 \cdot 010    \cdot $, then $c \rightarrow c+1 \Rightarrow  c=4$

\item $\overline S=1011\textcolor{red}{010}$, $\overline Q=1$, $\overline{SQ}=10110101$, $\overline{SQ}\pi=1011010$, $\overline{Q} \in v(\overline{SQ}\pi) $
$\rightarrow 1 \cdot 0 \cdot 11 \cdot 010 \cdot 1 \cdot $;

\item The end of $\overline{M}$ has been reached, 
so $c \rightarrow c+1 \Rightarrow c=5$. This leads to the 
partitioning $\overline{M} = 10110101 \Rightarrow 1 \cdot 0 \cdot 11 \cdot 010 \cdot 1  $ where $c=5$ is the number of partitions separated by dots. Beware that the length of the complete patterns follows the Fibonacci series: $1,1,2,3,\ldots$

\end{itemize}

\section{\label{sec:theta90}Scenarios for $\theta=\pi/2$: periodic jumps induces delocalization}

\begin{figure*}[h]
  \centering
\includegraphics[scale=0.9]{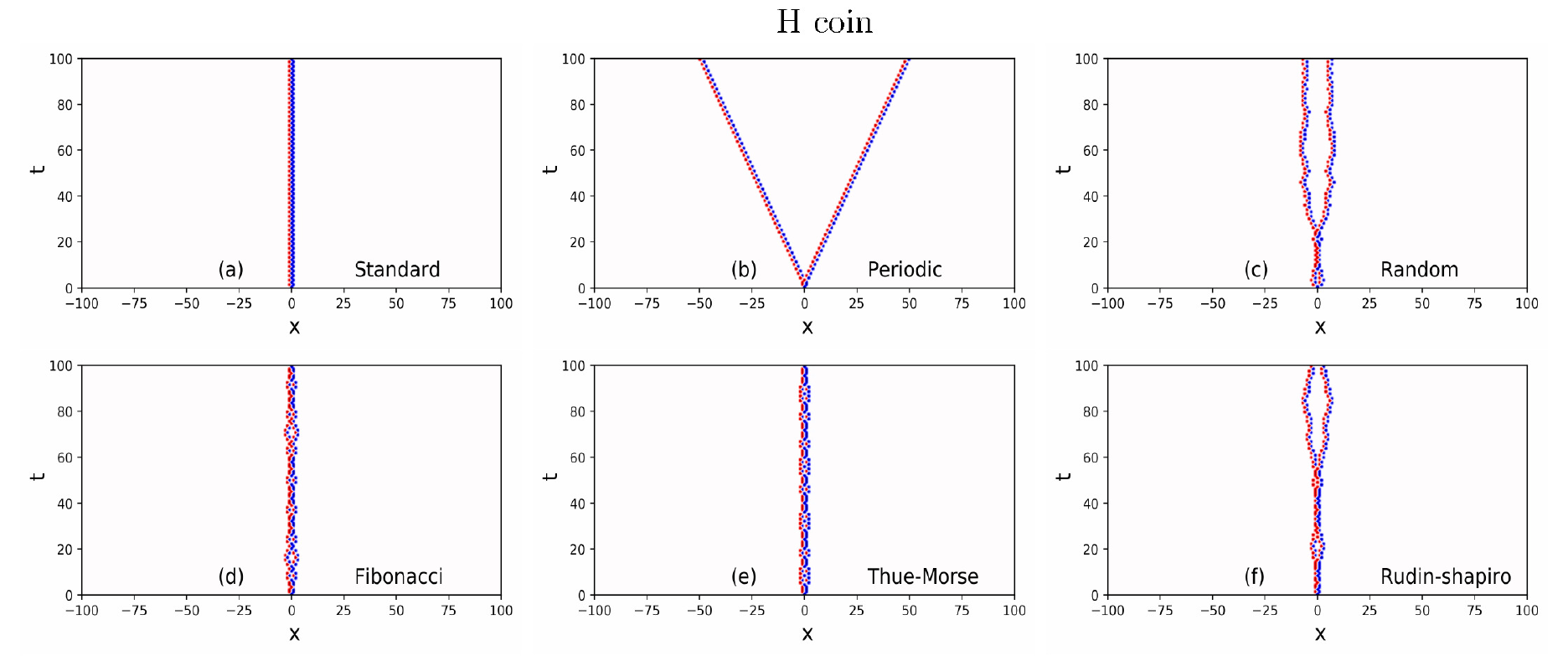}
\includegraphics[scale=0.9]{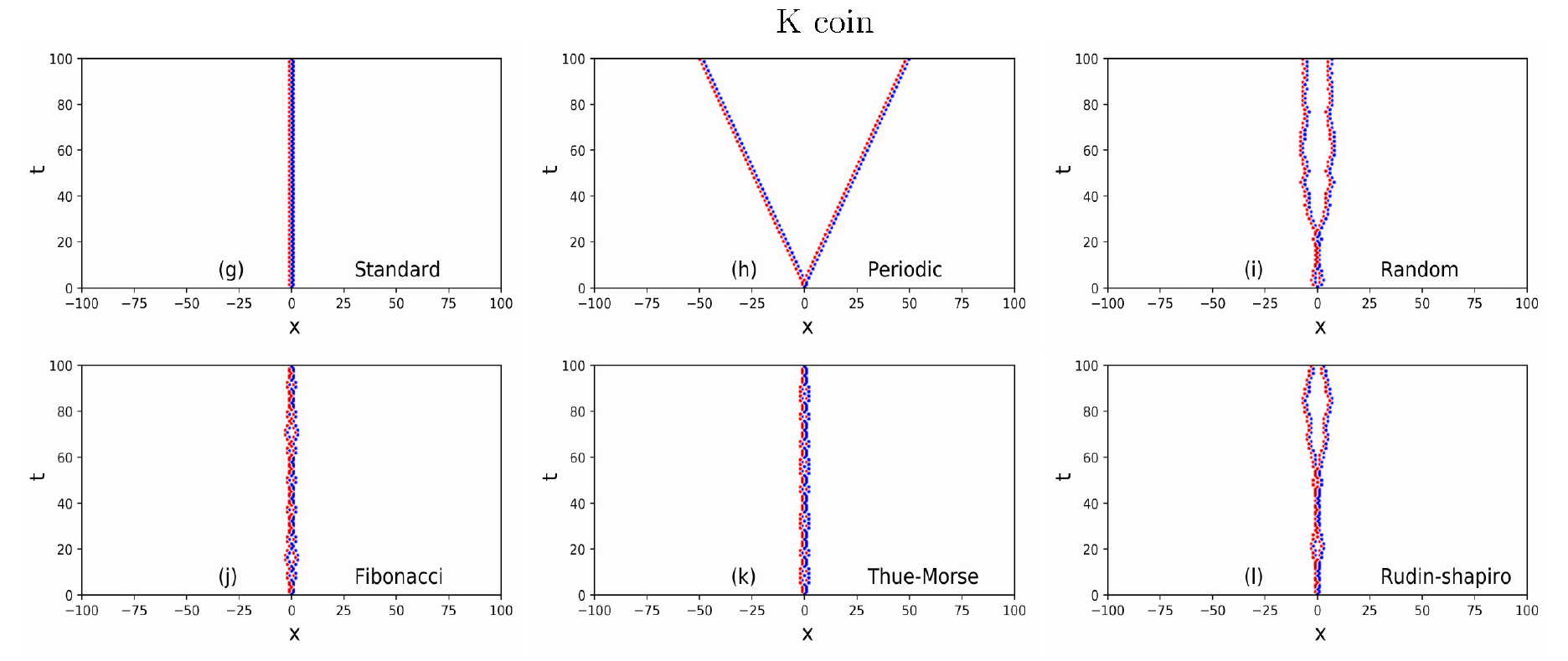}
\caption{Spatiotemporal evolution of the normalized asymmetry measure $A_t(x)/|A_t|^{\max}$ for  $\theta=\pi/2$. Quantum carpets for H coin (a-f) and K coin (g-l). Periodic jumps leads to delocalization.}
\label{fig:mxt90_hk}
\end{figure*}

In this appendix we provide an extra analysis of the spatiotemporal patterns for $\theta=\pi/2$ in Fig.\ref{fig:mxt90_hk}. In such scenarios the operators $\widehat C_{H,K}$ have pure nondiagonal terms since $c_{11}=c_{22}=0$.  The H coin becomes the X-Pauli matrices $\widehat C_{H}=\sigma_x$ that is also equivalent to the NOT-gate that acts flipping the the spin components:  $\ket{\downarrow}\rightarrow\ket{\uparrow}$ and  $\ket{\uparrow}\rightarrow\ket{\downarrow}$. The K coin does not precisely the NOT-gate, but it also acts flipping the spin components.  Such features of both coins lead to the alternating propagation-reflection effect: if the QW propagates at step $t$ then it will be reflected in the opposite orientation at $t+1$ and vice-versa. In turn, this forwards an oscillation around the initial position. Then it is straightforward to see that in such cases the QW will remain bounded localized near the origin. Indeed this happens for the Standard,  Fibonacci, Thue-Morse, Rudin-Shapiro and Random protocols. But interestingly, the periodic jumps boosts the escape from the fate of localization. 

\bibliography{qwbinaryjumps.bib}

\end{document}